\documentclass[10pt,conference]{IEEEtran}

\usepackage{cite}
\usepackage{graphicx}
\usepackage{url}
\usepackage{amsmath}
\usepackage{mathtools}
\usepackage{amsfonts}
\usepackage{multirow}
\usepackage[most]{tcolorbox}
\usepackage{xspace}
\usepackage{mdframed}
\usepackage{amssymb}
\usepackage{stmaryrd}
\usepackage{tabularx}
\usepackage{booktabs}
\usepackage{colortbl}
\usepackage[caption=false,font=footnotesize]{subfig}
\usepackage{tikz}
\usetikzlibrary{arrows.meta,positioning}

\newcommand{\toolname}{\textsc{Artic}\xspace}

\newcommand{\passkmetric}{{\fontencoding{T1}\selectfont\texttt{pass\textasciicircum k}}}

\begin{document}

\title{Natural-Language Workflows Are Not Software Yet: Artifact-Driven Compilation for Reliable Agent Execution}

\pagestyle{plain}
\pagenumbering{arabic}

\author{
\IEEEauthorblockN{
Xiangzhe Xu\IEEEauthorrefmark{1}, Hanxi Guo\IEEEauthorrefmark{1}, Guangyu Shen, Siyuan Cheng, Xiangyu Zhang}
\IEEEauthorblockA{
Purdue University\\
West Lafayette, USA\\
\{xzx, guo778, shen447, cheng535, xyzhang\}@purdue.edu}
\IEEEauthorrefmark{1}Equal contribution
}
\maketitle

\begin{abstract}
Natural-language workflows offer a software-like interface for agents: domain
experts can write reusable procedures, and agents can execute them as
instructions. This promise is not yet reliable. Workflow descriptions often
leave data dependencies implicit, so the executor must infer which prior
results a step should use; agents can also fail to follow long or branching
instructions under context pressure. We propose \toolname, an artifact-driven
workflow compiler that transforms a natural-language workflow into an
artifact-driven workflow in which each step declares the artifacts it reads and
writes, constraints gate produced artifacts, and explicit control transfers
route execution. This representation exposes the enforcement burden placed on
agent execution, allowing the compiler to identify steps that depend on too
much state or contain difficult control logic and refine them through
constrained optimization. To validate the LLM-assisted transformation,
\toolname decomposes faithfulness checking into local obligations and uses
scenario-based dry runs to test whether compiled workflow regions conform to
the source workflow. We evaluate \toolname on 488 problem instances from 11
real-world domain workflows; it improves task resolve rate by 28 percentage
points over the original text workflow. We also show that workflows compiled
by \toolname are 32 and 56 percentage points more consistent in cross-model
and repeated-execution setups, respectively.
\end{abstract}


\section{Introduction}

Natural-language workflows are emerging as a software-like layer for agents.
They encode intended behavior in natural language, while the agent runtime
interprets and executes that behavior~\cite{jiang2026aiware,hassan2024fmware,
chen2026promptware,liang2025prompts}. This software-like layer is attractive to domain experts.
Clinicians, analysts, and operations specialists can
specify how agents should perform domain work through skills, workflow prompts,
and procedural instruction templates~\cite{korinek2025ai,wang2026llm,
zhao2026ai,liu2026llm,hua2024trustagent,wang2026agentspec,kamath2025enforcing,
skills_oc,skills_hm,skills_claude,open_agent_spec}. In this view, a workflow
becomes a portable unit of domain expertise that can be shared, reused, and
executed by agents.

This software-like view carries a reliability expectation grounded in two properties of software. 
First, software can be
shared and reused across similar environments because its behavior is
predictable. Second, software execution is reliable because the machine is
expected to follow the program instructions exactly.

However, these properties are not yet realized for natural-language workflows due to the variability of how agents execute them.
Unlike
the underlying environment that runs conventional software, an agent may not
faithfully execute the natural-language workflow it receives.
The gap has two sources. First, natural-language workflows often leave data
dependencies implicit. Each step inherits all intermediate results in the past
context, resembling a program in which intermediate results are stored as
global variables: a later operation may read any of them even though the
dependency is not visible at the operation boundary. This implicit data flow
undermines executor reliability. At each step, an executor agent must
distinguish which prior results are relevant, and it may make mistakes when
the context contains noisy information.

Second, the effort of faithfully following workflow instructions grows with the workflow's control flow and context load.
Different
workflows impose different burdens on an agent. A short linear workflow may be
easy to follow when it fits within the agent's capability and context window; in contrast, a
workflow with many branches or heavy intermediate results in context is more
likely to cause skipped steps, wrong transfers, or lost information during
context compaction~\cite{liu2026plan,liu-etal-2024-lost,
jiang-etal-2024-followbench,wu-etal-2025-lifbench,shi2025flowagent,
chen2026agentif,purpura2026enhancing}.

The key challenge is to define an execution model for agentic workflows that
exposes their enforcement burden, so difficult workflow regions can be
identified and refined before runtime. Inspired by state-space views of
problem solving~\cite{newell1972human}, we formulate a natural-language
workflow as an explicit artifact-driven workflow.
Each step declares the artifacts it reads and writes, and explicit control
transfer conditions determine which step executes next. The artifact-driven workflow is
semi-structured: the local action of a step can still be a natural-language
instruction executed by an agent, but data dependencies and control transfers
are explicit and inspectable. This representation preserves agent flexibility
inside each step while giving the workflow concrete state and execution gates.

Since artifacts and control transfers are explicit, the artifact-driven workflow
thus exposes enforcement burden as concrete, analyzable signals.
For example, steps that depend on many artifacts require the agent to select relevant state
from a larger context and preserve it through context compaction. Likewise, nested or
compound routes increase the burden of following the intended control flow.
These signals identify which workflow regions are likely to be difficult for an
agent to execute faithfully.

We implement this idea as \toolname\footnote{\toolname: ARTIfact-driven workflow
Compiler.}, an artifact-driven workflow compiler, so domain experts do not
need to manually rewrite their workflows in this domain-specific language. Unlike
existing natural-language-to-workflow approaches that rely on unconstrained LLM decisions or repeated manual review~\cite{xu2024llm4workflow,liu2026masfactory,
zhong2026chat2workflow}, \toolname compiles a natural-language workflow into an artifact-driven form through constrained optimization: an LLM proposes and refines the artifact-driven workflow, while programmatic analyses over its explicit artifacts and control transfers provide feedback on context pressure and decision complexity.

The constraints used in the optimization above focus on reducing the execution
burden of the compiled workflow; they do not check whether the compiler LLM has
correctly translated the source workflow. In our ablation study, a compiler
variant without this correctness validation performs 16 percentage points worse
than the full system. It is therefore important to check whether the
artifact-driven workflow is source-consistent. This
check is challenging because, unlike a traditional compiler whose correctness
can be stated against formal source and target semantics~\cite{Leroy-Compcert-CACM},
\toolname transforms natural-language instructions whose semantics depend on
domain context and common-sense interpretation~\cite{ghosh2016arsenal,
cosler2023nl2spec,fuggitti2023nl2ltl}. We address this challenge by exploiting
an asymmetry in LLM systems: generating a source-consistent workflow is
difficult, but checking a smaller, concrete source-consistency claim can be
easier~\cite{cobbe2021training,lightman2023lets,madaan2023selfrefine}.

Concretely, the validator checks source consistency in two ways. First,
checking a whole workflow at once is unreliable, but the artifact-driven
workflow language lets the compiler decompose the validation goal into smaller
obligations that follow the language structure. Each local obligation is
simpler and easier to validate reliably. Second, following recent agent-testing
work~\cite{feng2025tai3,ahmed2026specops}, the compiler grounds validation in
concrete scenarios: it enumerates diverse cases and asks the LLM to dry-run
those cases to check whether the compiled workflow remains source-consistent
with the source workflow.

\noindent \textbf{Contributions.}
This paper makes four contributions.
\begin{itemize}
\setlength{\itemsep}{0pt}
\setlength{\parsep}{0pt}
  \item We formulate workflow enforcement as artifact-driven execution, exposing
  data dependencies, control transfers, and analyzable burden signals that
  indicate where faithful agent execution is at risk.
  \item We implement \toolname, a compiler prototype that transforms
  natural-language workflows into executable artifact-driven workflows through
  constrained optimization.
  \item We propose a validation paradigm for natural-language-to-workflow
  transformation, decomposing faithfulness checking into local obligations and
  scenario-based dry runs.
  \item We evaluate \toolname on 488 problem instances from 11 real-world
  domain workflows; it improves task resolve rate by 28 percentage points over
  the original text workflow. Workflows compiled by \toolname are also 32 and
  56 percentage points more consistent in cross-model and repeated-execution
  setups, respectively.
\end{itemize}

\section{Motivating Example}
\label{sec:motivation}

We use a patient-referral workflow derived from \mbox{\(\chi\)-Bench}~\cite{chibench2026},
a realistic healthcare benchmark for testing whether LLM agents can route
patients to the correct program. This example lets us examine why directly
executing a textual workflow is unreliable, why a naive executable workflow is
still insufficient, and how artifact-driven compilation motivates our design.

\begin{figure}[t]
\centering
\includegraphics[width=\linewidth]{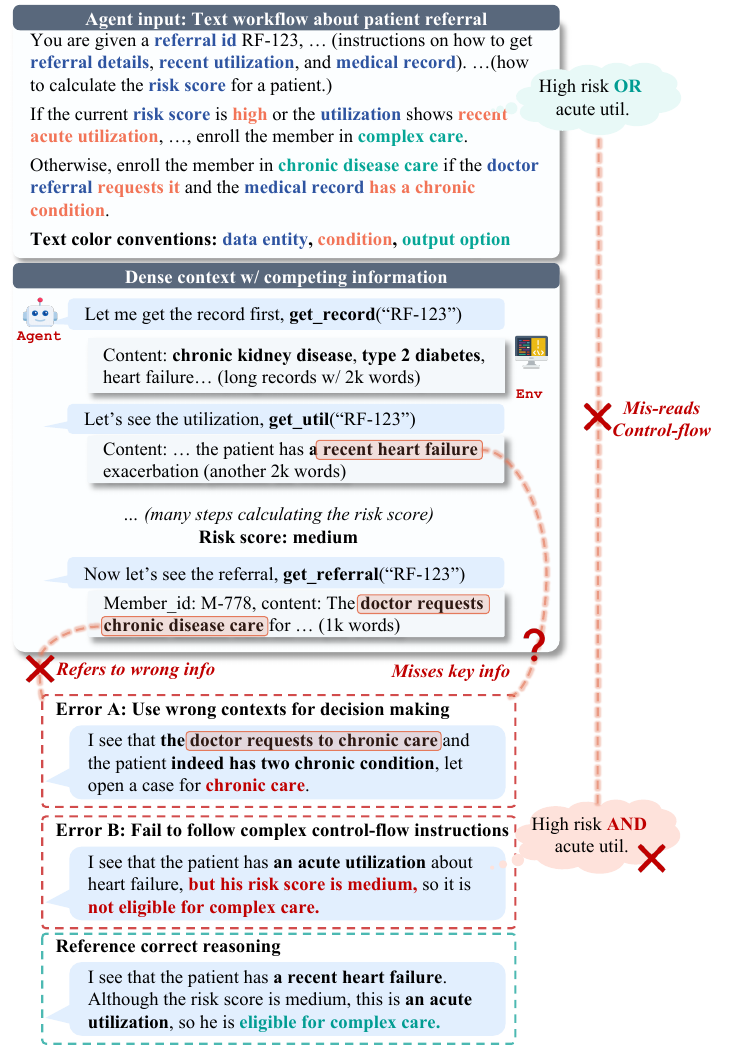}
\caption{Common failure modes when an agent follows a complex textual workflow.
The figure shows two potential failure modes together with the reference
correct reasoning. In Error A, the agent fails to select the correct
information from context for decision making: the key information is the recent
heart-failure exacerbation, which should support complex care, but the agent
mistakenly relies on the doctor's chronic-care referral message. In Error B,
the agent misreads a subtle control-flow instruction. The workflow specifies
high risk \emph{or} recent acute utilization, but the agent interprets it as
high risk \emph{and} recent acute utilization.}
\label{fig:motivating-text-workflow}
\end{figure}

\begin{figure*}[t]
\centering
\includegraphics[width=0.85\textwidth]{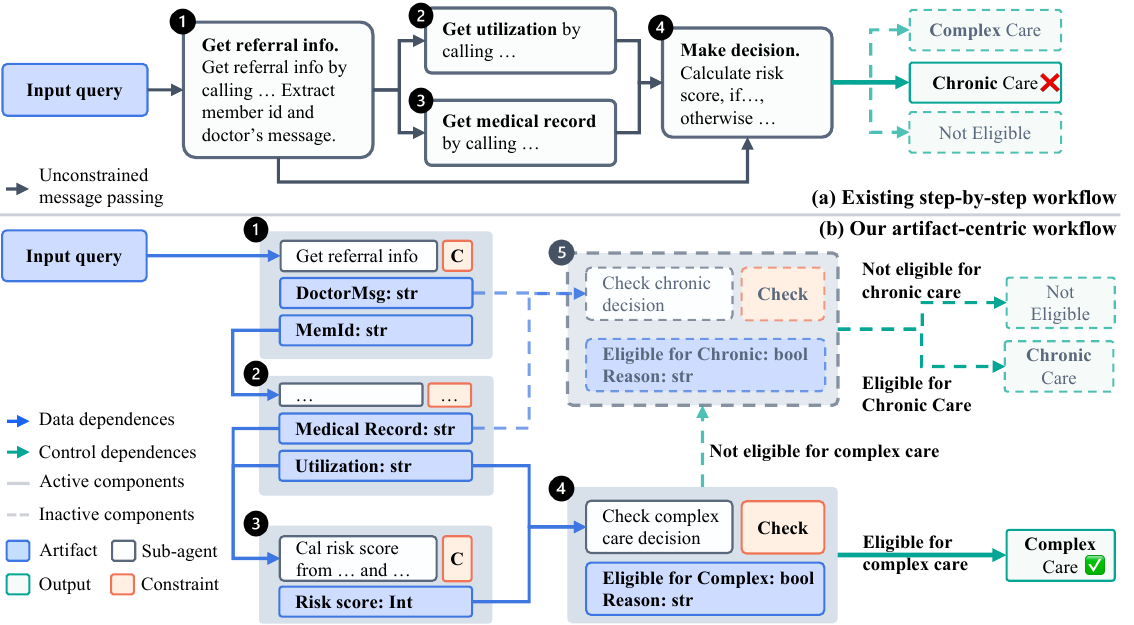}
\caption{Workflow structure affects enforcement reliability. (a)
shows a naive step-by-step workflow. The subagent in Step 4 still needs to
select the correct information from a dense context that includes the referral
message, medical record, utilization history, and other intermediate results.
Moreover, Step 4 still contains complex control-flow instructions that the
agent may misapply. (b) shows our artifact-driven workflow. The
compiler identifies the step that needs complex artifacts and decomposes it
into Step 3 and Step 4, so Step 4 only receives focused utilization information
and the risk score. The compiler also decomposes the complex control flow into
Step 4 and Step 5, so the workflow structure enforces the \emph{or} relation
specified in the original textual workflow.}
\label{fig:motivating-artifact-workflow}
\end{figure*}

\noindent
\textbf{Unstructured workflow text is not a reliable enforcement
mechanism.}
Figure~\ref{fig:motivating-text-workflow} shows the textual workflow setting.
The agent receives a referral id and retrieves the necessary medical data to
decide between complex care and chronic disease care. Error mode A illustrates
how a dense natural-language context can lead the agent to retrieve and use the
wrong information. The context contains the doctor's referral request,
medical-record evidence of chronic conditions, utilization notes about a recent
heart-failure exacerbation, and an intermediate risk assessment. The agent is
expected to use the utilization evidence to make the complex-care decision, but
because this dependency is not explicit, the executor agent may not consistently
identify the correct information. In the example, it instead attends to the
doctor's chronic-care request and chronic-condition evidence and makes the wrong
decision. Error mode B illustrates unreliable control-flow enforcement: the
complex-care route requires either a high risk score or recent acute
utilization. The agent may misread the subtle control-flow requirement,
mistakenly treating the condition as high risk \emph{and} recent acute
utilization, which leads to wrong instruction following.

\noindent \textbf{Naively converting textual workflows to executable form may
not improve enforcement.}
It is natural to translate a long textual workflow into several subagent calls
with separate contexts. The subagents then share information by sending
messages to the next agent, while deterministic Python code handles tool calls,
parsing, and routing. However, converting a natural-language workflow into this
step-by-step subagent-driven workflow does not necessarily improve enforcement.
Figure~\ref{fig:motivating-artifact-workflow} contrasts two executable
workflows for the same textual procedure. In the step-by-step workflow in
Figure~\ref{fig:motivating-artifact-workflow}(a), Step 4 still combines
referral information, utilization, medical records, and complex route
conditions. The subagent at that step still needs to select the correct data
from complex context, and the decision at that step still contains complex
control-flow structure.

\noindent \textbf{We enable more reliable execution by compiling workflows to
an artifact-driven language and adaptively decomposing challenging steps.}
To formulate the challenge introduced by complex context, we explicitly
introduce artifacts into the workflow: named intermediate results that a step
consumes or produces. In this representation, if a step accumulates or inputs
too many artifacts, it becomes an indicator of potential enforcement challenge.
Figure~\ref{fig:motivating-artifact-workflow}(b) shows an example. The
complex-care decision receives only the utilization and risk-score artifacts;
it does not receive the doctor's referral message, which is relevant to the
chronic-care decision. This separation reduces the chance that an agent selects
the wrong evidence from a long context. Furthermore, the artifact-driven
language lets the compiler measure control-flow complexity. In the example, the
compiler identifies that the original decision-making step combines risk-score
calculation, complex-care eligibility, and chronic-care eligibility. It
therefore decomposes the step into three simpler transformations: computing the
risk score, checking complex-care eligibility, and checking chronic-care
eligibility.

\section{Method}
\label{sec:method}

\begin{figure}[t]
\centering
\includegraphics[width=0.7\columnwidth]{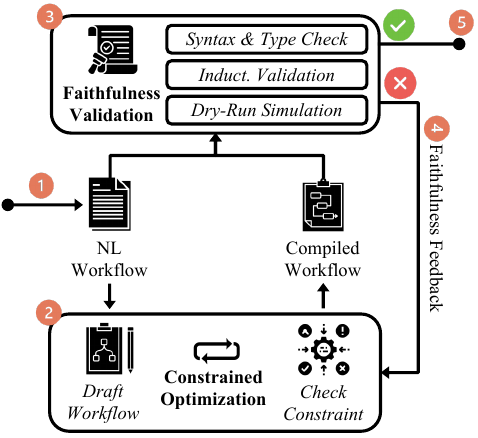}
\caption{Compiler workflow. The compiler drafts an artifact-driven workflow under
program-analysis feedback, validates faithfulness through static, inductive,
and dry-run checks, and either accepts the workflow or feeds diagnostics back to
the drafting stage.}
\label{fig:method-workflow}
\end{figure}

Figure~\ref{fig:method-workflow} gives the overall pipeline. The compiler has
two key designs. First, it formulates workflow compilation as constrained
optimization: the generated workflow should be easy for agents to enforce while
still preserving the source procedure. Second, it validates faithfulness before
accepting the generated workflow, because the transformation itself relies on
LLM reasoning and cannot be trusted from syntax alone.

Given an input natural-language workflow (step 1), the compiler asks an LLM to
compose a candidate artifact-driven workflow (step 2). The candidate isolates the
context needed by each step and decomposes steps that place too much reasoning
or context burden on one agent. This draft is guided by a constraint checker
that runs lightweight program analysis over the structured workflow and reports
suboptimal regions, such as excessive control-flow complexity or live artifacts
that would be hard to preserve through context compaction.

After a candidate workflow is drafted, a faithfulness-validation component
checks whether it preserves the intended behavior of the source workflow. The
validator has three passes. Static validation catches syntactic and
well-formedness errors. Inductive validation decomposes the global faithfulness
claim into smaller local validation goals over workflow regions. Dry-run
simulation helps discharge these local goals by simulating the execution of the
source and compiled regions on diverse hypothetical cases (step 3). If
validation finds an inconsistency, the compiler feeds the diagnostic back to the
constrained-optimization stage (step 4); otherwise, it accepts the workflow
(step 5).

\subsection{Artifact-Driven Workflow Formulation}

The core design of our language is to formulate a workflow as a sequence of
transformations over artifacts. We refer to this representation as an
artifact-driven workflow. This formulation lets the compiler explicitly measure
and analyze the enforcement challenge of a workflow step. To support this
discussion, we introduce a small language that formalizes artifact-driven
workflows. We illustrate the design by defining the semantics of this language;
the optimization and validation procedures in the following subsections are
also defined over the same language.

\noindent\textbf{Language definition.}
Figure~\ref{fig:artifact-driven-workflow-language-new} defines the core language. The four
statements denote the basic workflow structures. An agent step
\(\mathsf{agent}(\omega)\) asks a subagent to follow a natural-language
instruction \(\omega\) and complete a local artifact transformation. A
sequential step connects two consecutive workflow steps. An explicit branch
chooses between two subworkflows according to predicate \(p\), and a loop
repeats a subworkflow while \(p\) holds.

\begin{center}
\refstepcounter{figure}
\label{fig:artifact-driven-workflow-language-new}
\begin{mdframed}[
  linewidth=0.8pt,
  roundcorner=2pt,
  innerleftmargin=8pt,
  innerrightmargin=8pt,
  innertopmargin=6pt,
  innerbottommargin=6pt
]
\small
\[
\begin{array}{@{}r@{\;}c@{\;}l@{\qquad}l@{}}
S ::= 
  & \mathsf{agent}(\omega)
    && \text{agent step} \\[0.6mm]
  \mid & S_1 ; S_2
    && \text{sequential step} \\[0.6mm]
  \mid & \mathsf{if}\ p\ \mathsf{then}\ S_1\ \mathsf{else}\ S_2
    && \text{explicit branch} \\[0.6mm]
  \mid & \mathsf{while}\ p\ \mathsf{do}\ S
    && \text{loop}
\end{array}
\]
\end{mdframed}
{\footnotesize Fig.~\thefigure. Core artifact-driven workflow language.}
\end{center}

\noindent\textbf{Semantics definition.}
The semantics of the language is defined as transformations over artifact
states. We formally define an artifact state as a record that maps each
artifact identifier to its current data and semantic constraints:
\begin{equation}
\label{eq:new-artifact-state}
\Sigma : \mathsf{Aid} \rightharpoonup (\mathcal{D} \times 2^{\mathcal{C}}).
\end{equation}
Here, \(\mathsf{Aid}\) is the set of artifact identifiers, \(\mathcal{D}\) is
the domain of artifact data, and \(\mathcal{C}\) is the domain of semantic
constraints over artifacts. Thus \(\Sigma(i)=(d_i,C_i)\) says that artifact
\(i\) currently stores data \(d_i\) and is governed by the constraint set
\(C_i\). In the implementation, artifact data may be backed by a typed JSON
object, a document, or a file, while constraints record the semantic properties
that later checks and control transfers may inspect.

We further define two standard operations on artifact states. Reading a set of
artifact identifiers \(A\) restricts the store to those artifacts:
\begin{equation}
\label{eq:new-artifact-read}
\Sigma[A] = \{\,i \mapsto \Sigma(i) \mid i \in A\,\}.
\end{equation}
Updating an artifact state with produced artifact entries \(U\) overrides the
current data--constraint pair for the identifiers in \(U\):
\begin{equation}
\label{eq:new-artifact-update}
\Sigma \oplus U
=
\lambda i.
\begin{cases}
U(i), & i \in \mathsf{dom}(U),\\
\Sigma(i), & \text{otherwise.}
\end{cases}
\end{equation}

We denote the semantics of a statement with the following big-step notation:
\begin{equation}
\label{eq:new-big-step}
\langle S,\Sigma\rangle \Downarrow \Sigma'
\end{equation}
This means that statement \(S\), when executed from artifact state \(\Sigma\),
terminates in artifact state \(\Sigma'\). Figure~\ref{fig:main-semantics}
shows the core rules. In the agent rule, \(\mathsf{Run}_{\omega}(\Sigma)\)
denotes the possible artifact updates produced by executing instruction
\(\omega\) from state \(\Sigma\), and \(\mathsf{Check}_{\omega}(\Sigma,U)\)
denotes the constraint check over the produced update \(U\). The interpretation
of \textsc{Seq} and \textsc{If-True} is direct: sequential composition executes
the second statement from the state produced by the first, and the true branch
executes \(S_1\) when predicate \(p\) evaluates to true. The remaining branch
and loop rules are standard and are shown in
Figure~\ref{fig:supplemental-semantics} in
Appendix~\ref{app:validation-sketch}.

\begin{figure}[t]
\begin{mdframed}[linewidth=0.8pt,roundcorner=2pt]
\footnotesize
\setlength{\abovedisplayskip}{0.45em}
\setlength{\belowdisplayskip}{0.45em}

\begin{equation*}
\frac{
  U \in \mathsf{Run}_{\omega}(\Sigma)
  \qquad
  \mathsf{Check}_{\omega}(\Sigma,U)=\mathsf{ok}
}{
  \langle \mathsf{agent}(\omega),\Sigma\rangle
  \Downarrow
  \Sigma \oplus U
}
\tag{\textsc{Agent}}
\end{equation*}

\begin{equation*}
\frac{
  \langle S_1,\Sigma\rangle \Downarrow \Sigma_1
  \qquad
  \langle S_2,\Sigma_1\rangle \Downarrow \Sigma_2
}{
  \langle S_1;S_2,\Sigma\rangle \Downarrow \Sigma_2
}
\tag{\textsc{Seq}}
\end{equation*}

\begin{equation*}
\frac{
  \llbracket p\rrbracket(\Sigma)=\mathsf{true}
  \qquad
  \langle S_1,\Sigma\rangle \Downarrow \Sigma'
}{
  \langle \mathsf{if}\ p\ \mathsf{then}\ S_1\ \mathsf{else}\ S_2,\Sigma\rangle
  \Downarrow \Sigma'
}
\tag{\textsc{If-True}}
\end{equation*}
\end{mdframed}
\caption{Core big-step semantics for artifact-driven workflows.}
\label{fig:main-semantics}
\end{figure}

\subsection{Optimization Objectives}

We formulate workflow compilation as an iterative constrained optimization
process. The compiler uses a set of optimization objectives to guide the LLM
toward a workflow that is easier to enforce while preserving the source
procedure. The intuition follows the challenges illustrated in
Section~\ref{sec:motivation}: if a natural-language step contains too much
context or too much control structure, an agent may struggle to follow it
faithfully. The compiler should therefore keep only enforceable instructions in
one step and decompose complex instructions into multiple steps.

Estimating context and control complexity is nontrivial because the structure
is latent in a prompt. The artifact-driven language makes this analysis
possible by normalizing a prompt into a workflow and then applying program
analysis to that workflow. Given a prompt \(\omega\), we write
\(\mathsf{fg}(\omega)\) for its finest-grained artifact-driven workflow representation. In
this view, each agent step has one local goal and produces one focused
artifact. This fine-grained view exposes the latent context and control-flow
structure that the original prompt would otherwise ask one agent to manage
implicitly. We use four optimization objectives to formulate this constrained
optimization problem.

\noindent\textbf{Control-flow complexity.}
The intuition of control-flow complexity is that a prompt with many logical
branches is harder for an agent to enforce faithfully. We denote the
control-flow burden of a prompt \(\omega\) as \(b_{\mathsf{cf}}(\omega)\) and
estimate it by applying cyclomatic complexity to its finest-grained workflow:
\begin{equation}
\label{eq:new-branch-complexity}
b_{\mathsf{cf}}(\omega)
=
\widehat{\mathsf{cyclo}}(\mathsf{fg}(\omega)).
\end{equation}
Here \(\widehat{\mathsf{cyclo}}\) is an estimate of the cyclomatic complexity
of the prompt's latent workflow.

\noindent\textbf{Context pressure.}
The intuition of context pressure is that a prompt is harder to enforce when
one agent invocation must carry a long execution context before reaching a
checkpoint boundary. We first define a context estimator \(C\) over a
fine-grained workflow. Let
\(\widehat{\mathsf{ctx}}(X)\) be an atomic estimate of the context length for a
primitive body or guard \(X\):
\begin{equation}
\label{eq:new-context-pressure}
\begin{aligned}
C(\mathsf{agent}(\omega))
&=
\widehat{\mathsf{ctx}}(\omega)
\\
C(S_1;S_2)
&=
C(S_1)+C(S_2)
\\
C(\mathsf{if}\ p\ \mathsf{then}\ S_1\ \mathsf{else}\ S_2)
&=
\max\!\left(\widehat{\mathsf{ctx}}(p), C(S_1), C(S_2)\right)
\\
C(\mathsf{while}\ p\ \mathsf{do}\ S)
&=
\widehat{k}(p,S)\cdot C(S).
\end{aligned}
\end{equation}
The sequential case adds pressure because the same prompt asks one agent to
carry both pieces of latent work. Branches take the larger route because only
one route executes. Loops scale by the estimated iteration count
\(\widehat{k}\). We denote the context burden of a prompt \(\omega\) as
\(b_{\mathsf{ctx}}(\omega)\):
\begin{equation}
\label{eq:new-agent-context-pressure}
b_{\mathsf{ctx}}(\omega)=C(\mathsf{fg}(\omega)).
\end{equation}

\begin{figure}[t]
\centering
\includegraphics[width=0.85\columnwidth]{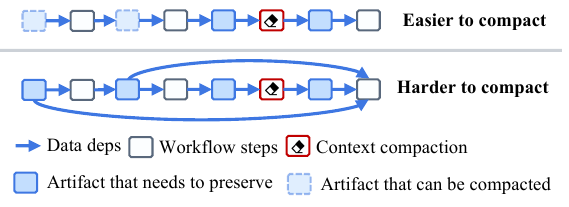}
\caption{Context-compaction pressure. A step is harder to compact when many
live artifacts must be preserved across the compaction point.}
\label{fig:opt-ctx-pressure}
\end{figure}

\noindent\textbf{Compaction pressure.}
The intuition of compaction pressure is that a prompt is harder to execute
robustly when important information must survive a context-compaction point.
For a fine-grained workflow \(G\), assume \(\mathsf{Live}_{G}(\ell)\) is
provided by standard live-variable analysis at program point \(\ell\). The
pointwise compaction pressure is:
\begin{equation}
\label{eq:new-pointwise-compaction-pressure}
p_{\mathsf{cmp}}(G,\ell)
=
\sum_{a\in\mathsf{Live}_{G}(\ell)} s(a).
\end{equation}
We denote the compaction burden of a prompt \(\omega\) as
\(b_{\mathsf{cmp}}(\omega)\) and use the worst internal compaction point:
\begin{equation}
\label{eq:new-agent-compaction-pressure}
b_{\mathsf{cmp}}(\omega)
=
\max_{\ell\in\mathsf{pts}(\mathsf{fg}(\omega))}
p_{\mathsf{cmp}}(\mathsf{fg}(\omega),\ell).
\end{equation}
Figure~\ref{fig:opt-ctx-pressure} illustrates this case: when many live
artifacts cross a compaction point, the compactor must preserve more state for
future execution to remain well-defined.

\noindent\textbf{Artifact cost.}
Artifact cost tracks the overhead introduced by materializing intermediate
results. We denote the artifact cost of a workflow \(W\) as \(A(W)\):
\begin{equation}
\label{eq:new-artifact-cost}
A(W)=\sum_{a\in\mathsf{Art}(W)} s(a),
\end{equation}
where \(\mathsf{Art}(W)\) is the set of materialized artifacts in workflow
\(W\), and \(s(a)\) is the estimated size of artifact \(a\). The compiler
therefore searches for workflows that reduce local control-flow, context, and
compaction pressure without introducing unnecessary artifact materialization.

\subsection{Faithfulness Validation}

The optimization step relies on LLM reasoning, so its edits can involve
nontrivial natural-language judgments. The compiler must therefore check
whether the produced workflow is a faithful representation of the source
workflow, but this check also requires LLM reasoning. The key challenge is to
make validation easier than generation, so that an LLM validator with comparable
capability can still improve the compiler. This generation-validation asymmetry
has been exploited in verifier, process-supervision, and self-feedback methods
for LLM reasoning~\cite{cobbe2021training,lightman2023lets,madaan2023selfrefine}.

We reduce the challenge of validation by two key ideas. First, validating a whole workflow at once is
difficult, but the artifact-driven workflow language lets the compiler decompose the
validation goal into smaller obligations that follow the language structure.
Each local obligation is simpler and easier to validate reliably. Second, the
compiler facilitates LLM reasoning by concretizing the context: it enumerates
diverse concrete test cases and asks the LLM to check whether the compiled
workflow remains faithful when simulating those cases.

\noindent\textbf{Static well-formedness.}
Before invoking LLM validators, the compiler applies deterministic static
checks to the artifact-driven workflow program. These checks validate the surface syntax,
ensure that every artifact read is declared and available, and type-check
artifact availability at control-flow joins. The key type check is path
stability at joins. If a program point can be reached from multiple
predecessors, then every incoming path must provide compatible types for the
artifacts read at that point:
\begin{equation}
\label{eq:wf-join-types}
\begin{aligned}
&\forall \ell.\ \forall p_1,p_2\in\mathsf{Pred}(\ell).\\
&\qquad
\mathsf{type}_{p_1}(\mathsf{Reads}(\ell))
\simeq
\mathsf{type}_{p_2}(\mathsf{Reads}(\ell)).
\end{aligned}
\end{equation}
Here, \(\mathsf{Pred}(\ell)\) is the set of predecessor paths that can reach
program point \(\ell\), \(\mathsf{Reads}(\ell)\) is the set of artifacts read
at \(\ell\), and \(\mathsf{type}_{p}(\cdot)\) gives the artifact types
available along path \(p\).

This check helps the compiler make latent execution obligations explicit. A
natural-language workflow may describe only the condition under which an
artifact is normally produced, while a later step implicitly assumes that the
artifact is available. For example, suppose a workflow says: ``if the patient
has a high risk score, retrieve the patient's medical record and report any
chronic disease,'' followed by ``cross-check the patient's utilization.'' A
verbatim branch interpretation leaves the medical record unavailable on the
low-risk path before the second step, even though the intended cross-check may
still require that record. The join type check exposes this mismatch: either
the workflow must retrieve a compatible medical-record artifact on all paths,
or the later step must be rewritten so that its declared inputs match the
artifacts actually available.

\begin{figure*}[t]
\centering
\includegraphics[width=0.8\textwidth]{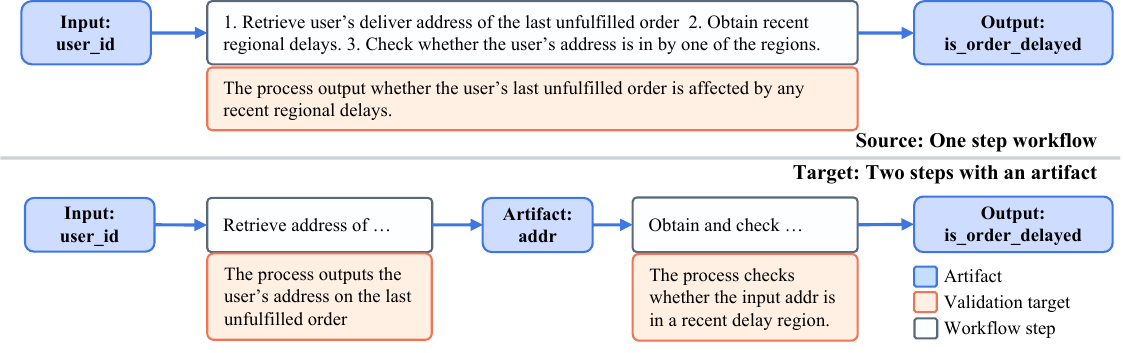}
\caption{How we validate workflow by induction. Suppose the source workflow
describes one natural-language step, but the compiled workflow refines it into
two subworkflows \(S_1;S_2\) connected by an intermediate artifact. The
validator first checks that the artifact is a valid conceptual boundary in the
source workflow: after \(S_1\), the artifact should contain exactly the
intermediate state that the remaining source workflow may rely on. It then
validates the prefix up to that artifact and validates the continuation from
that artifact. If both local obligations hold, the sequence validation composes
them into a validation of the full compiled fragment.}
\label{fig:valid-induct}
\end{figure*}

\noindent\textbf{Inductive validation.}
The compiler validates by induction over the workflow program structure. It
decomposes the global validation goal into smaller obligations that match the
structure of the compiled program, and asks the validator to discharge each
obligation separately. Figure~\ref{fig:valid-induct} illustrates this
decomposition for a sequence, where a single validation goal is reduced to an
artifact-boundary check and two local validation obligations.
Appendix~\ref{app:faithfulness-validation-sketch} gives the formal proof sketch
for this validation structure.

\begin{figure}[t]
\centering
\includegraphics[width=0.75\linewidth]{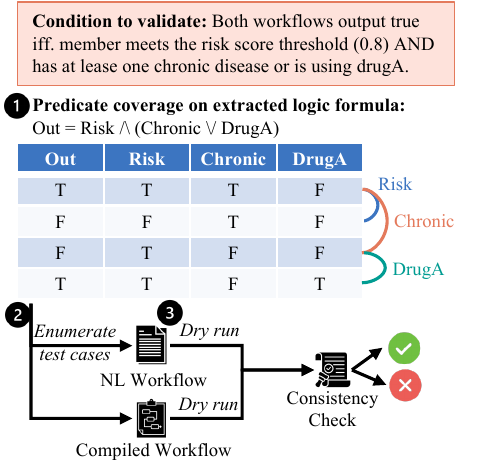}
\caption{Dry-run simulation validation. The compiler extracts predicates,
enumerates coverage-guided scenarios, and compares independent dry runs of the
source and compiled workflows.}
\label{fig:dry-run-validation}
\end{figure}

\noindent\textbf{Dry-run simulation.}
The compiler further facilitates faithfulness validation by enumerating
concrete test scenarios and asking LLM validators to reason over those
scenarios. Scenario-based testing has been shown effective for exposing agent
misinterpretation and execution failures~\cite{feng2025tai3,ahmed2026specops}.
Our dry-run simulation has three stages. First, it extracts a logical formula
from the validation target and identifies the atomic predicates that should be
tested. Second, it uses predicate coverage to enumerate concrete scenarios,
including cases where changing one predicate changes the expected artifact or
route. Third, it asks LLM validators to independently simulate the source and
compiled workflows on each scenario, and compares the simulated trajectories
and outcomes to detect inconsistencies. Figure~\ref{fig:dry-run-validation}
illustrates the process.

\section{Implementation}
\label{sec:implementation}

We implement the compiler prototype core in 5.9K lines of Python code. The code and data will be available upon publication.
The resulting executable workflow package has
three parts. The \emph{orchestrator} is a deterministic Python script that
invokes individual statements and transfers control according to each step's
output. A \emph{step} delegates local natural-language instructions to
subagents and parses their produced artifacts. The \emph{runtime} implements
low-level interfaces, including resolving concrete artifact paths from
artifact identifiers, managing artifact versions, and invoking subagents.
Appendix~\ref{app:implementation-details} gives additional implementation
details.

\section{Evaluation}
\label{sec:evaluation}
\definecolor{paperblue100}{HTML}{E6ECF6}
\definecolor{paperblue200}{HTML}{C9D5EB}
\definecolor{papercoral500}{HTML}{F36F52}

\subsection{Setup}

\noindent\textbf{Benchmarks.}
Our workflow compiler is intended to help non-programming experts build more
reliable agentic workflows for domain-specific procedures. We therefore
evaluate on 11 domains drawn from SOP-Bench~\cite{sopbench2026} and
\(\chi\)-Bench~\cite{chibench2026}. SOP-Bench contains standard operating
procedures (SOPs) designed to reflect real-world industrial workflows, with
executable interfaces and human-validated test cases. \(\chi\)-Bench is a realistic
healthcare workflow simulator with MCP tools, medical-operation handbooks, and
multi-role interaction. Rather than running the full end-to-end healthcare
process, we use its care-management intake and program-eligibility stage,
which aligns with our goal of evaluating complex domain-workflow enforcement.
We report this domain as \emph{Medical}. For domains with more than 50
instances, we sample at most 50 for budget reasons. In total, our test set
contains 488 instances from 11 domains; detailed domain statistics appear in
Table~\ref{tab:domain-workflow-stats}.

\noindent\textbf{Execution setup.}
We evaluate three compiler models: GPT-5.4, Sonnet-4.6, and GLM-5. These are
frontier large models used to translate textual workflows into artifact-driven
workflows. We also evaluate six executor models, ranging from 3B active
parameters to 700B+ total parameters among models with public size metadata.
These executors are cost-efficient enough for realistic workflows in
real-world domains. The full model list appears in Appendix
Table~\ref{tab:evaluation-models}. All executions use the same Codex-based
agent harness, because Codex provides a general tool-executing interface and
has been used beyond software engineering for knowledge-work tasks such as
document drafting, data analysis, and communication
coordination~\cite{openai_codex_docs,johnston2026codex}.
The evaluated agent interacts with each benchmark environment through its MCP
tools. The primary metric is task resolve rate: the fraction of benchmark
instances whose final output satisfies the benchmark oracle. These benchmarks
are designed to test workflow following: an instance is resolved only when the
agent follows the prescribed procedure and makes the correct domain decision.
We compare against textual execution, text-space skill rewriting, direct code
generation, and existing
natural-language-to-workflow baselines.

\begin{table*}[t]
\centering
\caption{Comparison of effectiveness across different workflow
representations. Numbers denote task resolve rate in percent.}
\label{tab:effectiveness-by-domain-model}
\begingroup
\definecolor{oursbg}{HTML}{E6ECF6}
\newcommand{\oursval}[1]{\cellcolor{oursbg}#1}
\newcommand{\best}[1]{\textbf{#1}}
\footnotesize
\renewcommand{\arraystretch}{1.08}
\begin{tabular*}{\textwidth}{@{\extracolsep{\fill}}lcccccccccccc}
\toprule
\multirow{2}{*}{Domain}
  & \multicolumn{4}{c}{GLM-4.7-Flash}
  & \multicolumn{4}{c}{GPT-OSS-120B}
  & \multicolumn{4}{c}{Qwen3-235B} \\
\cmidrule(lr){2-5}\cmidrule(lr){6-9}\cmidrule(lr){10-13}
  & Text & SkillCreator & Code & \toolname
  & Text & SkillCreator & Code & \toolname
  & Text & SkillCreator & Code & \toolname \\
\midrule
Medical
  & 48 & 72 & 12 & \oursval{\best{88}}
  & 36 & 72 & 12 & \oursval{\best{76}}
  & 40 & 56 & 8 & \oursval{\best{76}} \\
Customer service
  & 52 & 72 & 48 & \oursval{\best{88}}
  & 24 & 34 & 48 & \oursval{\best{66}}
  & 64 & 90 & 48 & \oursval{\best{92}} \\
Dangerous goods
  & 38 & 84 & 94 & \oursval{\best{98}}
  & 68 & 90 & 94 & \oursval{\best{98}}
  & 54 & 90 & 94 & \oursval{\best{98}} \\
Know your business
  & 48 & 52 & \best{66} & \oursval{64}
  & 44 & 32 & \best{66} & \oursval{62}
  & 48 & 52 & \best{66} & \oursval{62} \\
Order fulfillment
  & 86 & 83 & \best{90} & \oursval{\best{90}}
  & 0 & 0 & \best{90} & \oursval{\best{90}}
  & \best{90} & \best{90} & \best{90} & \oursval{\best{90}} \\
Patient intake
  & 85 & 91 & \best{100} & \oursval{\best{100}}
  & 9 & 12 & \best{100} & \oursval{97}
  & 94 & \best{100} & \best{100} & \oursval{\best{100}} \\
Referral abuse v1
  & 84 & 80 & \best{100} & \oursval{98}
  & 92 & 74 & \best{100} & \oursval{\best{100}}
  & 54 & 56 & \best{100} & \oursval{\best{100}} \\
Referral abuse v2
  & 50 & 36 & 82 & \oursval{\best{96}}
  & 76 & 66 & 82 & \oursval{\best{100}}
  & 36 & 64 & 82 & \oursval{\best{100}} \\
Traffic spoofing
  & 68 & 66 & 66 & \oursval{\best{74}}
  & 46 & 36 & 66 & \oursval{\best{74}}
  & 72 & \best{74} & 66 & \oursval{\best{74}} \\
Video annotation
  & 72 & 34 & \best{90} & \oursval{\best{90}}
  & 50 & 40 & \best{90} & \oursval{\best{90}}
  & 32 & 86 & \best{90} & \oursval{\best{90}} \\
Warehouse inspection
  & 52 & \best{58} & 50 & \oursval{50}
  & \best{58} & 48 & 50 & \oursval{50}
  & 80 & \best{82} & 50 & \oursval{50} \\
\midrule
Average
  & 62 & 66 & 73 & \oursval{\best{85}}
  & 46 & 46 & 73 & \oursval{\best{82}}
  & 60 & 76 & 72 & \oursval{\best{85}} \\
\bottomrule
\end{tabular*}
\endgroup
\end{table*}

\noindent\textbf{Research questions.}
The evaluation is organized around four research questions. RQ1 evaluates
whether artifact-driven compilation improves workflow enforcement. RQ2
evaluates the reliability of the compiled workflows under model variation,
repeated execution, and execution stressors. RQ3 analyzes compilation time and
cost. RQ4 ablates how each system component contributes to system
performance. We then present two case studies
to illustrate how dry-run simulation improves compilation and where the current
pipeline remains limited.

\subsection{RQ1: Does Artifact-Driven Compilation Improve Workflow
Enforcement?}

Table~\ref{tab:effectiveness-by-domain-model} compares artifact-driven
workflows with representative baselines from two families. The first family
relies on the general capability of LLM agents to improve the workflow.
\emph{SkillCreator} takes the original workflow as input and rewrites it into a
clearer skill document, but the result is still textual instructions rather
than an executable workflow. \emph{Code} instructs a coding agent to write
Python code with LLM calls that implements the workflow, using the same harness
and executor model as our compiler. We provide this baseline with the necessary Python
wrappers for MCP tools and utility functions for invoking subagents. Besides,
\emph{Text} denotes the setup that directly gives the original workflow to the
executor, and \toolname denotes the compiled artifact-driven workflow.

Overall, our compiled workflows achieve the best average task resolve rate for
every executor model: 85\% on GLM-4.7-Flash, 82\% on GPT-OSS-120B, and 85\%
on Qwen3-235B. This indicates that artifact-driven compilation improves
workflow enforcement beyond simply rewriting the instruction or asking an
agent to implement it as code. On domains such as patient intake and referral
abuse, the code baseline can slightly outperform our workflow while our
workflow remains near-perfect. Inspection suggests that these code
implementations often use stable heuristics such as keyword matching, which can
work well for simple domain rules but may not generalize to complex workflows,
as reflected by the suboptimal performance of \emph{Code} on other domains. For
domains such as Know Your Business and warehouse inspection, all
representations have relatively low performance. These workflows contain
properties that require model judgment, such as distinguishing an ordinary typo
from fake business information. This challenge is orthogonal to workflow
enforcement; Figure~\ref{fig:kyb-typo-case-study} gives a concrete example.

The second baseline family consists of specially designed pipelines that
convert natural language into executable workflows~\cite{liu2026masfactory,
zhong2026chat2workflow}.
Table~\ref{tab:nl2workflow-baselines} compares our compiler with two such
baselines on GLM-4.7-Flash. MASFactory often fails before producing an
executable workflow on these benchmark domains, and Chat2Workflow can produce
runnable workflows for some domains but expects more concrete human guidance
about the intended workflow structure. It is less accurate when workflows
become more complex. These results suggest that existing techniques lack an
autonomous validation and feedback mechanism that can ensure workflow quality
in complex domains with limited human involvement.

\begin{table}[t]
\centering
\caption{Comparison between our compiler and existing NL-to-workflow baselines. ERR denotes that the baseline failed to
generate an executable workflow.}
\label{tab:nl2workflow-baselines}
\begingroup
\definecolor{oursbg}{HTML}{E6ECF6}
\newcommand{\oursval}[1]{\cellcolor{oursbg}#1}
\newcommand{\best}[1]{\textbf{#1}}
\newcommand{\err}{\textsc{err}}
\footnotesize
\renewcommand{\arraystretch}{1.08}
\begin{tabular*}{\columnwidth}{@{\extracolsep{\fill}}lccc}
\toprule
Domain & MASFactory & Chat2Workflow & \toolname \\
\midrule
Medical & 0 & 52 & \oursval{\best{88}} \\
Customer service & 0 & 0 & \oursval{\best{88}} \\
Dangerous goods & \err & 0 & \oursval{\best{98}} \\
Know your business & \err & 0 & \oursval{\best{64}} \\
Order fulfillment & 0 & 0 & \oursval{\best{90}} \\
Patient intake & 0 & \best{100} & \oursval{\best{100}} \\
Referral abuse v1 & \err & \best{100} & \oursval{98} \\
Referral abuse v2 & 0 & \best{100} & \oursval{96} \\
Traffic spoofing & \err & 0 & \oursval{\best{74}} \\
Video annotation & \err & 0 & \oursval{\best{90}} \\
Warehouse inspection & \err & 0 & \oursval{\best{50}} \\
\midrule
Average & 0 & 32 & \oursval{\best{85}} \\
\bottomrule
\end{tabular*}
\endgroup
\end{table}

\begin{table}[t]
\centering
\caption{Domain-level workflow statistics.}
\label{tab:domain-workflow-stats}
\scriptsize
\setlength{\tabcolsep}{2.5pt}
\begin{tabular*}{\columnwidth}{@{\extracolsep{\fill}}@{}lrrrrr@{}}
\toprule
Domain & Cases & Length & Nodes & Edges & Artifacts \\
\midrule
Medical & 25 & 22,133 & 8 & 13 & 8 \\
Customer service & 50 & 9,166 & 10 & 18 & 12 \\
Dangerous goods & 50 & 5,638 & 9 & 11 & 8 \\
Know your business & 50 & 8,772 & 15 & 19 & 11 \\
Order fulfillment & 29 & 2,631 & 6 & 6 & 6 \\
Patient intake & 34 & 4,848 & 7 & 7 & 7 \\
Referral abuse v1 & 50 & 8,277 & 6 & 6 & 6 \\
Referral abuse v2 & 50 & 12,558 & 12 & 14 & 10 \\
Traffic spoofing & 50 & 4,954 & 8 & 8 & 8 \\
Video annotation & 50 & 31,431 & 14 & 20 & 14 \\
Warehouse inspection & 50 & 4,292 & 8 & 9 & 11 \\
\midrule
\textbf{Average} & 44.4 & 10,427 & 9.4 & 11.9 & 9.2 \\
\bottomrule
\end{tabular*}
\end{table}

\subsection{RQ2: Are Compiled Workflows More Reliable?}

\begin{figure}[t]
\centering
\includegraphics[width=.9\columnwidth]{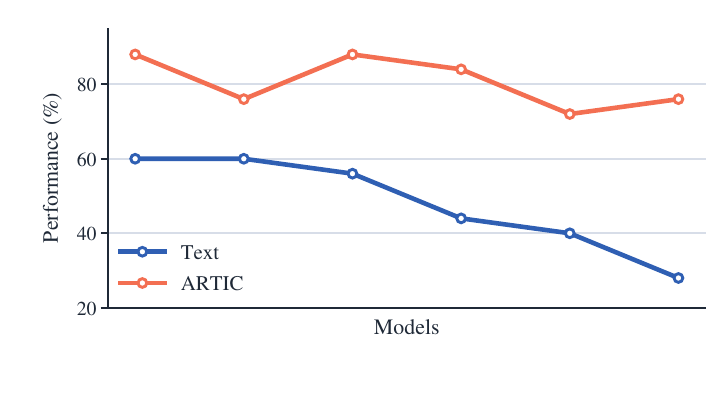}
\caption{Performance across executor models on the Medical domain.}
\label{fig:chi-model-size-performance}
\vspace{-10pt}
\end{figure}

\noindent\textbf{Consistency across executor models.}
The results show that compiled workflows behave more stably when executed by
different executor models. Figure~\ref{fig:chi-model-size-performance}
compares the Medical domain from \(\chi\)-Bench across six executor models. The
text workflow varies from 28\% on its worst model to 60\% on its best model,
twice the spread of the compiled workflow, which ranges from 72\% to 88\%.
The smaller degradation suggests that explicit artifacts and control transfers
make execution less sensitive to the executor model.

\begin{figure}[t]
\centering
\includegraphics[width=\columnwidth]{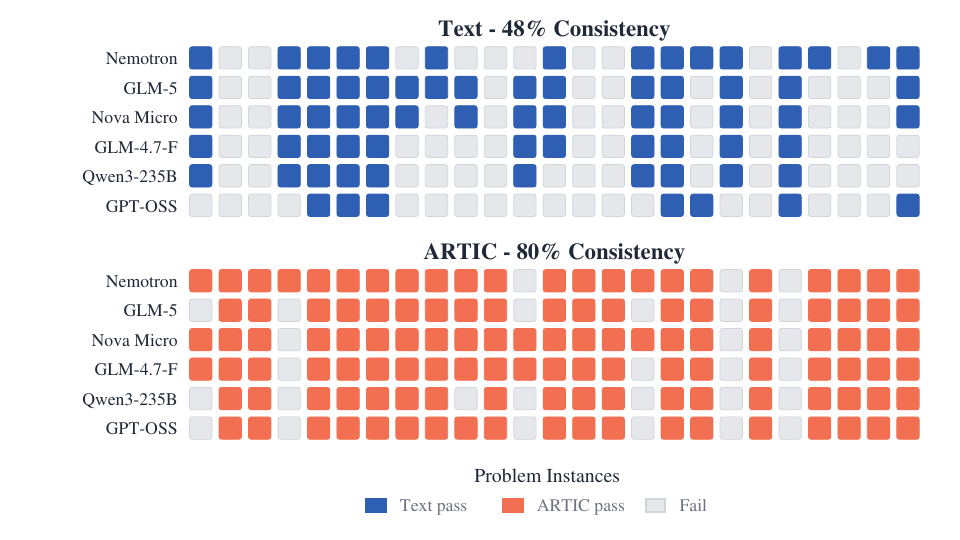}
\caption{Per-instance consistency across executor models on the Medical domain.
Each row is a model and each block is one problem instance; colored blocks pass
and gray blocks fail.}
\label{fig:chi-model-size-consistency}
\end{figure}

\noindent\textbf{Per-instance consistency.}
Besides overall performance variance, it is also important to measure
per-instance consistency. We call an instance consistent when all six executor
models either pass it or fail it under the same workflow representation.
Suppose a user develops and debugs a workflow on one executor model, observes
which cases succeed or fail, and later deploys the same workflow with a
different executor. This metric measures how often that development-time
observation transfers across executor environments.
Figure~\ref{fig:chi-model-size-consistency} shows pass/fail outcomes for the
same Medical-domain instances across executor models. The compiled workflow is
consistent on 80\% of instances, compared with 48\% for the text baseline.

\begin{figure}[t]
\centering
\includegraphics[width=.9\columnwidth]{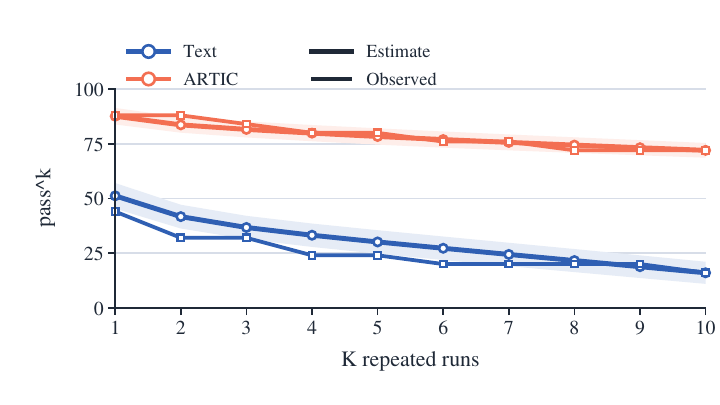}
\caption{Consistency of workflows in repeated runs. The metric \passkmetric{}
denotes the probability that all \(k\) repeated runs are correct. Solid and
dashed lines show the unbiased estimator and observed performance,
respectively.
Shaded bands show \(\pm\) one standard deviation.}
\label{fig:chi-stability-passk}
\end{figure}

\noindent\textbf{Consistency across repeated executions.}
We use \passkmetric, following Tau-Bench~\cite{yao2024taubench}, to measure
how likely all \(k\) repeated runs are correct when the same workflow is
executed by the same model multiple times. Figure~\ref{fig:chi-stability-passk}
shows the result on the Medical domain in \(\chi\)-Bench. We report both the
unbiased estimator and the observed performance; Appendix~\ref{app:passk-details}
gives the calculation details. At \(k=10\), the compiled workflow has about
72\% passing cases, while the text workflow has only 16\%. From \(k=1\) to
\(k=10\), the compiled workflow degrades by 16 percentage points, compared
with 35 points for the text workflow. This means the compiled workflow is
significantly more stable than the text workflow under repeated execution.

\noindent\textbf{Robustness under environment perturbations.}
We test the robustness of each workflow under environment perturbations. In
real-world execution, an agent may encounter transient model request errors or
context compaction. This test measures how well a workflow tolerates such
perturbations. We use all data samples where both the text and compiled
workflow give the correct answer in the original setting, then observe how
their performance changes under different conditions. The compiled workflow
preserves 100\% performance, while the text workflow drops to 80\% under the
strongest perturbations. See Figure~\ref{fig:chi-robustness-radar} in the appendix for details.

\noindent\textbf{Per-agent context pressure.}
We further evaluate how compiled workflows reduce context burden by decomposing
challenging steps into smaller subagents. Across the
Table~\ref{tab:effectiveness-by-domain-model} samples, the compiled workflow
reduces average per-agent input tokens by 63\% and output tokens by 50\%
compared with an agent working on the text workflow.
Figure~\ref{fig:context-pressure-token-plots} in the appendix gives the full
box plots by executor model.

\subsection{RQ3: What Is the Compilation Cost?}

Compilation is an offline cost paid once per workflow. For the studied
SOP-Bench domains, the median first-package compilation time with GLM-5 is
under 5 minutes, and the average compilation cost is under \$3.
Figure~\ref{fig:sop-compilation-time-cost} in the appendix gives the full
time-and-cost distribution.

\subsection{RQ4: How Does Each System Component Contribute?}

We ablate the dry-run feedback, validation-by-induction, and optimization
components of our system. Each component contributes meaningfully: the observed
performance changes range from 8 to 32 percentage points. Optimization
contributes the most, which is expected because it identifies when a prompt
places too much context or decision burden on the executor and decomposes that
step. We also test different compiler models and find that the generated
workflow performance is stable across compiler choices. Detailed ablation and
compiler-model results appear in Appendix~\ref{app:additional-evaluation-results}.

\subsection{Case Studies}

\begin{figure*}[t]
\centering
\includegraphics[width=.83\textwidth]{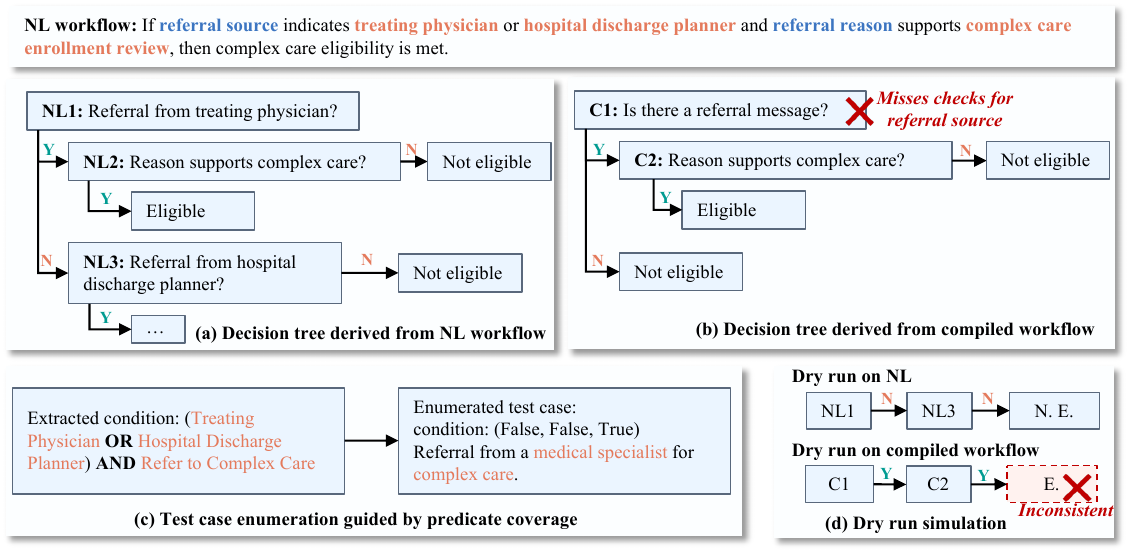}
\caption{Dry-run feedback case study. The dry run compares the intended
natural-language condition with the compiled decision structure on a
representative case, identifying missed checks for referral source before
runtime evaluation.}
\vspace{-10pt}
\label{fig:case-study-dryrun}
\end{figure*}

\noindent\textbf{How dry-run simulation helps validate compiled workflows.}
Figure~\ref{fig:case-study-dryrun} shows a concrete case in which dry-run
simulation identifies an inconsistency in a compiled workflow. The compiled
workflow used an overly vague prompt for checking the source of a referral:
as shown in Figure~\ref{fig:case-study-dryrun}(b), the first decision node
checked whether a referral message exists, but did not verify whether the
referral came from a qualifying source. The dry-run process begins with test
case enumeration guided by predicate coverage, as shown in
Figure~\ref{fig:case-study-dryrun}(c). The exposing case is a referral from a
non-qualifying person with a reason that otherwise supports complex-care
review. Figure~\ref{fig:case-study-dryrun}(d) then runs two independent
simulations: one over the source natural-language workflow and one over the
compiled workflow. Neither dry run is given the oracle label. Instead, each
simulation follows its workflow step by step. The resulting traces disagree,
which localizes the error to the referral-source predicate and provides the
compiler with targeted feedback.

\noindent\textbf{Limitation of workflow compiler.}
The workflow compiler does not achieve perfect accuracy. Cases that depend
more on model judgment than workflow enforcement may remain unresolved.
Figure~\ref{fig:kyb-typo-case-study} illustrates a concrete example in the
``Know your business'' domain, where the SOP asks the executor to decide
whether mismatches among a business name, website, address, and email are
ordinary typos or evidence of fabricated information. Artifact-driven
compilation can make the relevant inputs explicit and route the decision
through a focused step, but the final judgment still depends on the executor
model's knowledge and calibration. This challenge is orthogonal to workflow
enforcement.

\begin{figure}[t]
\centering
\begin{tcolorbox}[
  enhanced,
  width=\columnwidth,
  colback=paperblue100,
  colframe=paperblue200,
  fontupper=\footnotesize,
  boxrule=0.6pt,
  arc=1pt,
  left=5pt,
  right=5pt,
  top=5pt,
  bottom=5pt
]
\textbf{Original text from the SOP of ``Know your business''.}
The SOP first instructs the agent to obtain the website, address, and email
address of the business; this paragraph explains how to decide whether further
information is needed. ``They should use their experience to check if the
website and/or address and/or email matches with the pre-validated business
name. Due to errors by associates, there may be typos. \ldots{}
{\color{papercoral500}\textbf{\textit{Use your experience to identify if the
errors are genuine typos by associates entering the data or made-up names
submitted by the businesses.}}} If you believe that
there is some irregularities here, please enter in the system
\texttt{awaiting information}.''
\end{tcolorbox}
\caption{KYB example in which the source workflow explicitly requires judgment
about whether a mismatch is an ordinary typo or a data
irregularity.}
\label{fig:kyb-typo-case-study}
\end{figure}

\section{Related Work}

\noindent \textbf{Specification languages and workflow enforcement.}
Software-engineering research has long studied specification languages for
making expected behavior precise~\cite{jackson2002alloy,abrial2010modeling,van2003workflow,van2005yawl,pesic2007declare}. Our goal is different from asking users to write such a
specification language directly. Domain experts often start from
natural-language workflows, and the hard step is to translate those workflows
into an enforceable representation. Existing specification languages typically
still rely on humans to decompose a procedure, choose the right abstraction
boundaries, and encode the intended control and data dependencies~\cite{parnas1972criteria,jackson1995world,zave1997four}. This paper
instead studies how a compiler can recover an artifact-driven representation
from natural language and then make the result checkable and
enforceable by an agent runtime.

\noindent \textbf{Automatic skill generation and prompt optimization.}
Another line of work generates skills, prompts, or agent policies from task
feedback\cite{zhou2022large,pryzant2023automatic,madaan2023selfrefine,shinn2023reflexion,wang2023voyager,khattab2023dspy,zelikman2022star}. This direction is largely orthogonal to our
goal. It is especially useful when the task is outcome driven and the system
can freely search for a better prompt or strategy, as in test-driven software
development. Domain workflows have a different status: they encode expert
requirements about the procedure itself, so the compiler cannot arbitrarily
change the workflow to improve outcome metrics. Our problem is therefore not
only to obtain a better-performing prompt, but to preserve and enforce the
workflow constraints that the expert intended.

\noindent \textbf{Skill compilers and skill adaptation.}
Concurrent work on skill compilation studies how agent systems can retrieve
relevant skills and adapt skills across different harnesses~\cite{xu2026skillsmith,ouyang2026skcc,chen2026skillcraft,li2026single,ma2026skillgen}. That layer is complementary to ours. Skill
retrieval decides which skill should be used, and harness adaptation rewrites a
skill into a semantically equivalent counterpart for a different execution
environment. This paper focuses on a different question: once a domain
workflow is selected, how can the runtime measure and enforce whether the agent
faithfully follows it? Our artifact-driven representation makes enforcement
itself an object of analysis through explicit artifacts, constraints, and
control transfers.

\noindent \textbf{Natural language to workflow generation.}
Existing work around natural-language-to-workflow generation falls into three
related streams. First, some systems synthesize or evolve new workflows and
rules to improve task performance~\cite{li2024autoflow,gurkan2025lear}. This
direction is orthogonal to ours: expert workflows are requirements to be
preserved, not search spaces that the system can freely rewrite. Second,
systems such as Chat2Workflow and MASFactory help users construct executable
workflows, but typically rely on human steering or correction to reach a valid
workflow~\cite{zhong2026chat2workflow,liu2026masfactory}. Our setting instead
asks the compiler to validate and refine the workflow with limited manual
intervention. Third, platform-oriented systems and benchmarks operate in
settings where component semantics are more fixed, such as workflow
orchestration platforms or enterprise workflow services~\cite{fan2025workflowllm,liu2025workteam,xiao2024flowbench,xu2024llm4workflow}. These systems provide useful
front-end techniques, but the fixed platform semantics may not generalize to other domains.
\toolname focuses on that enforcement problem: the generated workflow must make
intermediate artifacts, control transfers, and validation obligations explicit
enough for the runtime to check and enforce.

\section{Conclusion}
\label{sec:conclusion}

Natural-language workflows are easy to write but hard to enforce. This paper
formulates them as artifact-driven workflows, where artifacts expose state,
constraints gate progress, and control transfers are explicit. \toolname
compiles natural-language workflows into this form and improves enforcement
across realistic domains. Its validation combines well-formedness checks,
compositional reasoning, and scenario-based dry runs.

\bibliographystyle{IEEEtran}
\bibliography{main}

\clearpage
\appendices
\section{Validation Details}
\label{app:validation-sketch}

\subsection{Faithfulness Validation Formal Sketch}
\label{app:faithfulness-validation-sketch}

This subsection gives the formal sketch behind the faithfulness-validation
discussion in Section~\ref{sec:method}. Let \(W_{\mathsf{n}}\) denote a source
natural-language workflow fragment, and let \(W_{\mathsf{c}}\) denote the
compiled artifact-driven workflow fragment. The validation goal is not to prove
syntactic equivalence between the two representations. Instead, it asks whether
the compiled workflow preserves the observable artifact behavior of the source
workflow under dry-run simulation.

Figure~\ref{fig:supplemental-semantics} gives the branch and loop rules omitted
from the main text. They use the same artifact-state judgment as
Equation~\ref{eq:new-big-step}.

\noindent\textbf{Observational correspondence.}
We first define when two artifact stores agree on the artifacts that matter to
the outside workflow:
\begin{equation}
\label{eq:app-obs-store-correspondence}
\begin{aligned}
&\Sigma_{\mathsf{n}}\sim_{\mathsf{obs}}\Sigma_{\mathsf{c}}
\coloneqq\\
&\quad
\forall a\in\mathsf{Obs}(\Sigma_{\mathsf{n}}).\
\exists b\in\mathsf{Obs}(\Sigma_{\mathsf{c}}).\
a\simeq b
\\
&\quad{}\land
\forall b\in\mathsf{Obs}(\Sigma_{\mathsf{c}}).\
\exists a\in\mathsf{Obs}(\Sigma_{\mathsf{n}}).\
a\simeq b.
\end{aligned}
\end{equation}
Here, \(\mathsf{Obs}(\Sigma)\) is the set of observable artifacts in store
\(\Sigma\), including final outputs, exposed route decisions, and intermediate
artifacts whose values affect later observable behavior. The relation
\(a\simeq b\) means that artifacts \(a\) and \(b\) are semantically equivalent
for the workflow: they may differ in surface wording, but they agree on the
decisions, required fields, and constraint-relevant content.

The correspondence is intentionally one-directional in the validation goal
below: compiled workflow behavior should be allowed by the source workflow, but
the compiled workflow need not reproduce every possible interpretation a reader
might infer from underspecified natural-language text. The comparison also
tracks only observable artifacts. Runtime scratch artifacts, provenance records,
and repair feedback may differ as long as they affect observable behavior only
through declared artifact interfaces.

\noindent\textbf{Dry-run big-step semantics.}
Because the source workflow is written in natural language, validation uses an
LLM policy \(\pi\) as a dry-run simulator. Given workflow \(W\) and artifact
store \(\Sigma\), the policy samples a rollout
\(r \sim \pi(\cdot \mid W,\Sigma)\). The rollout records a simulated
step-by-step execution plan, including the artifacts produced, decisions made,
and rationale used by the simulator. A deterministic interpretation function
\(\mathsf{Apply}\) updates the artifact store according to the rollout. We
write this as the dry-run big-step judgment:
\begin{equation}
\label{eq:app-dry-run-big-step}
\frac{
  r \sim \pi(\cdot \mid W,\Sigma)
  \qquad
  \Sigma'=\mathsf{Apply}(r,\Sigma)
}{
  \langle W,\Sigma\rangle \Downarrow_{\pi} \Sigma'
}.
\end{equation}
The judgment \(\langle W,\Sigma\rangle \Downarrow_{\pi} \Sigma'\) therefore
means that policy \(\pi\), when asked to dry run workflow \(W\) from input
artifact store \(\Sigma\), produces a rollout whose artifact effects yield
\(\Sigma'\).

\begin{figure}[t]
\begin{mdframed}[linewidth=0.8pt,roundcorner=2pt]
\footnotesize
\setlength{\abovedisplayskip}{0.45em}
\setlength{\belowdisplayskip}{0.45em}

\begin{equation*}
\frac{
  \llbracket p\rrbracket(\Sigma)=\mathsf{false}
  \qquad
  \langle S_2,\Sigma\rangle \Downarrow \Sigma'
}{
  \langle \mathsf{if}\ p\ \mathsf{then}\ S_1\ \mathsf{else}\ S_2,\Sigma\rangle
  \Downarrow \Sigma'
}
\tag{\textsc{If-False}}
\end{equation*}

\begin{equation*}
\frac{
  \llbracket p\rrbracket(\Sigma)=\mathsf{false}
}{
  \langle \mathsf{while}\ p\ \mathsf{do}\ S,\Sigma\rangle
  \Downarrow \Sigma
}
\tag{\textsc{While-False}}
\end{equation*}

\begin{equation*}
\frac{
  \llbracket p\rrbracket(\Sigma)=\mathsf{true}
  \qquad
  \langle S,\Sigma\rangle \Downarrow \Sigma_1
  \qquad
  \langle \mathsf{while}\ p\ \mathsf{do}\ S,\Sigma_1\rangle
  \Downarrow \Sigma_2
}{
  \langle \mathsf{while}\ p\ \mathsf{do}\ S,\Sigma\rangle
  \Downarrow \Sigma_2
}
\tag{\textsc{While-True}}
\end{equation*}
\end{mdframed}
\caption{Supplemental big-step semantic rules.}
\label{fig:supplemental-semantics}
\end{figure}

\begin{tcolorbox}[
  title={Goal. Dry-run observational preservation},
  colback=white,
  colframe=black,
  fonttitle=\bfseries,
  boxrule=0.6pt,
  arc=1pt,
  left=4pt,
  right=4pt,
  top=4pt,
  bottom=4pt
]
The validation goal is observational refinement: every observable behavior
produced by the compiled workflow should be allowed by the source
natural-language workflow under the dry-run policy. Formally, we write:
\begin{equation}
\label{eq:app-observational-refinement}
\begin{aligned}
&W_{\mathsf{c}}\preceq_{\mathsf{obs}} W_{\mathsf{n}}
\coloneqq
\forall \Sigma_{\mathsf{n}},\Sigma_{\mathsf{c}},
\Sigma'_{\mathsf{c}}.\\
&\Sigma_{\mathsf{n}}\sim_{\mathsf{obs}}\Sigma_{\mathsf{c}}
\ \land\
\langle W_{\mathsf{c}},\Sigma_{\mathsf{c}}\rangle
\Downarrow_{\pi}\Sigma'_{\mathsf{c}}
\\
&\Rightarrow
\exists \Sigma'_{\mathsf{n}}.\\
&\langle W_{\mathsf{n}},\Sigma_{\mathsf{n}}\rangle
\Downarrow_{\pi}\Sigma'_{\mathsf{n}}
\ \land\
\Sigma'_{\mathsf{n}}\sim_{\mathsf{obs}}\Sigma'_{\mathsf{c}}.
\end{aligned}
\end{equation}
\end{tcolorbox}

Here, \(W_{\mathsf{c}}\preceq_{\mathsf{obs}} W_{\mathsf{n}}\) means that every
observable behavior of the compiled workflow is allowed by the source
natural-language workflow under the dry-run policy. This is the property
approximated by the dry-run validation procedure in the main text. The
test-case generator selects representative input stores
\((\Sigma_{\mathsf{n}},\Sigma_{\mathsf{c}})\), including corner cases guided by
predicate coverage. The paired dry runs instantiate the two
\(\Downarrow_{\pi}\) judgments. The consistency checker then tests whether the
resulting stores satisfy \(\sim_{\mathsf{obs}}\).

\smallskip
\noindent\textbf{Initial.}
The base case is the trivial compiled form: an atomic agent step whose
instruction is exactly the source natural-language fragment. This rule has no
premise because it is definitional.
\begin{equation}
\label{eq:app-val-initial}
\frac{}{
\mathsf{agent}(W_{\mathsf{n}})\preceq_{\mathsf{obs}} W_{\mathsf{n}}
}
\tag{\textsc{Val-Initial}}
\end{equation}

\smallskip
\noindent\textbf{Sequential.}
The sequential induction step simultaneously checks a source decomposition and
composes the two compiled refinements. The source-decomposition premise says
that if the original region \(W_{\mathsf{n}}\) dry runs from \(\Sigma\) to
\(\Sigma_2\), and the proposed source pieces \(W_{\mathsf{n},1}\) and
\(W_{\mathsf{n},2}\) dry run through an intermediate state \(\Sigma_1\), then
their final state \(\Sigma'_2\) must match \(\Sigma_2\):
\begin{equation}
\label{eq:app-seq-source-decomposition}
\begin{aligned}
&\mathsf{Decomp}(W_{\mathsf{n}},W_{\mathsf{n},1},W_{\mathsf{n},2})\\
&\quad\coloneqq
\forall \Sigma,\Sigma_1,\Sigma_2,\Sigma'_2.\\
&\quad
\langle W_{\mathsf{n}},\Sigma\rangle\Downarrow_{\pi}\Sigma_2
\ \land\
\langle W_{\mathsf{n},1},\Sigma\rangle\Downarrow_{\pi}\Sigma_1
\ \land\
\langle W_{\mathsf{n},2},\Sigma_1\rangle\Downarrow_{\pi}\Sigma'_2
\\
&\quad\Rightarrow
\Sigma'_2\sim_{\mathsf{obs}}\Sigma_2 .
\end{aligned}
\end{equation}
Under this decomposition condition, refinements of the two pieces imply
refinement of the original region:
\begin{equation}
\label{eq:app-val-seq}
\frac{
\begin{aligned}
&W_{\mathsf{c},1}\preceq_{\mathsf{obs}}W_{\mathsf{n},1}
\qquad
W_{\mathsf{c},2}\preceq_{\mathsf{obs}}W_{\mathsf{n},2}
\\
&\mathsf{Decomp}(W_{\mathsf{n}},W_{\mathsf{n},1},W_{\mathsf{n},2})
\end{aligned}
}{
W_{\mathsf{c},1};W_{\mathsf{c},2}
\preceq_{\mathsf{obs}}
W_{\mathsf{n}}
}
\tag{\textsc{Val-Seq}}
\end{equation}
\emph{Proof.} Suppose the compiled sequence runs from a compiled input
\(\Sigma_{\mathsf{c}}\) that matches the source input \(\Sigma_{\mathsf{n}}\).
Formally, assume:
\begin{equation}
\label{eq:app-val-seq-proof-assume}
\begin{aligned}
&\Sigma_{\mathsf{n}}\sim_{\mathsf{obs}}\Sigma_{\mathsf{c}},\\
&\langle W_{\mathsf{c},1},\Sigma_{\mathsf{c}}\rangle
\Downarrow_{\pi}\Sigma_{\mathsf{c},1},\\
&\langle W_{\mathsf{c},2},\Sigma_{\mathsf{c},1}\rangle
\Downarrow_{\pi}\Sigma_{\mathsf{c},2}.
\end{aligned}
\end{equation}
We must construct a source output that matches \(\Sigma_{\mathsf{c},2}\). By
\(W_{\mathsf{c},1}\preceq_{\mathsf{obs}}W_{\mathsf{n},1}\) and the first two
lines of Equation~\ref{eq:app-val-seq-proof-assume}, there exists a source
intermediate state \(\Sigma_1\) such that:
\begin{equation}
\label{eq:app-val-seq-proof-prefix}
\langle W_{\mathsf{n},1},\Sigma_{\mathsf{n}}\rangle
\Downarrow_{\pi}\Sigma_1
\quad\land\quad
\Sigma_1\sim_{\mathsf{obs}}\Sigma_{\mathsf{c},1}.
\end{equation}
Then, by
\(W_{\mathsf{c},2}\preceq_{\mathsf{obs}}W_{\mathsf{n},2}\), Equation~\ref{eq:app-val-seq-proof-prefix},
and the last line of Equation~\ref{eq:app-val-seq-proof-assume}, there exists a
state \(\Sigma'_2\) such that:
\begin{equation}
\label{eq:app-val-seq-proof-suffix}
\langle W_{\mathsf{n},2},\Sigma_1\rangle
\Downarrow_{\pi}\Sigma'_2
\quad\land\quad
\Sigma'_2\sim_{\mathsf{obs}}\Sigma_{\mathsf{c},2}.
\end{equation}
Finally, let \(\Sigma_2\) be the dry-run result of the original source region:
\begin{equation}
\label{eq:app-val-seq-proof-source}
\langle W_{\mathsf{n}},\Sigma_{\mathsf{n}}\rangle
\Downarrow_{\pi}\Sigma_2.
\end{equation}
Applying the source-decomposition condition from
Equation~\ref{eq:app-seq-source-decomposition} to
Equations~\ref{eq:app-val-seq-proof-prefix},
\ref{eq:app-val-seq-proof-suffix}, and~\ref{eq:app-val-seq-proof-source} gives:
\begin{equation}
\label{eq:app-val-seq-proof-match-source}
\Sigma'_2\sim_{\mathsf{obs}}\Sigma_2.
\end{equation}
By transitivity of \(\sim_{\mathsf{obs}}\), Equations~\ref{eq:app-val-seq-proof-suffix}
and~\ref{eq:app-val-seq-proof-match-source} imply
\(\Sigma_2\sim_{\mathsf{obs}}\Sigma_{\mathsf{c},2}\). Thus \(\Sigma_2\) is the
witness required by the refinement definition in
Equation~\ref{eq:app-observational-refinement}. \hfill\(\square\)

\smallskip
\noindent\textbf{Branch.}
The branch induction step checks that guard \(\chi\) partitions the source
region into the intended true and false cases. The source case-split premise
says that, on states where the guard evaluates to true, every behavior of the
true source fragment is allowed by the original source region; on states where
the guard evaluates to false, every behavior of the false source fragment is
allowed by the original source region:
\begin{equation}
\label{eq:app-branch-source-split}
\begin{aligned}
&\mathsf{Split}_{\chi}(
W_{\mathsf{n}},W_{\mathsf{n},t},W_{\mathsf{n},f})\\
&\quad\coloneqq
\forall \Sigma,\Sigma'.\\
&\quad
\left(
\begin{aligned}
&\llbracket \chi \rrbracket(\Sigma)=\mathsf{true}
\ \land\
\langle W_{\mathsf{n},t},\Sigma\rangle
\Downarrow_{\pi}\Sigma'\\
&\Rightarrow
\exists \widehat{\Sigma}'.
\langle W_{\mathsf{n}},\Sigma\rangle
\Downarrow_{\pi}\widehat{\Sigma}'
\land
\widehat{\Sigma}'\sim_{\mathsf{obs}}\Sigma'
\end{aligned}
\right)
\\
&\quad{}\land
\left(
\begin{aligned}
&\llbracket \chi \rrbracket(\Sigma)=\mathsf{false}
\ \land\
\langle W_{\mathsf{n},f},\Sigma\rangle
\Downarrow_{\pi}\Sigma'\\
&\Rightarrow
\exists \widehat{\Sigma}'.
\langle W_{\mathsf{n}},\Sigma\rangle
\Downarrow_{\pi}\widehat{\Sigma}'
\land
\widehat{\Sigma}'\sim_{\mathsf{obs}}\Sigma'
\end{aligned}
\right).
\end{aligned}
\end{equation}
The guard-consistency premise says that the guard has the same value on
observationally corresponding source and compiled states:
\begin{equation}
\label{eq:app-branch-guard-consistency}
\begin{aligned}
\mathsf{Consistent}_{\chi}
\coloneqq
\forall \Sigma_{\mathsf{n}},\Sigma_{\mathsf{c}}.\
&\Sigma_{\mathsf{n}}\sim_{\mathsf{obs}}\Sigma_{\mathsf{c}}
\Rightarrow\\
&\llbracket \chi \rrbracket(\Sigma_{\mathsf{n}})
=
\llbracket \chi \rrbracket(\Sigma_{\mathsf{c}}).
\end{aligned}
\end{equation}
Under the case-split and guard-consistency conditions, refinements of the two
source cases imply that the compiled branch refines the original source region:
\begin{equation}
\label{eq:app-val-branch}
\frac{
\begin{aligned}
&W_{\mathsf{c},t}\preceq_{\mathsf{obs}}W_{\mathsf{n},t}
\qquad
W_{\mathsf{c},f}\preceq_{\mathsf{obs}}W_{\mathsf{n},f}
\\
&\mathsf{Split}_{\chi}(W_{\mathsf{n}},W_{\mathsf{n},t},W_{\mathsf{n},f})
\\
&\mathsf{Consistent}_{\chi}
\end{aligned}
}{
\mathsf{if}\ \chi\ \mathsf{then}\ W_{\mathsf{c},t}\ \mathsf{else}\ W_{\mathsf{c},f}
\preceq_{\mathsf{obs}}
W_{\mathsf{n}}
}
\tag{\textsc{Val-Branch}}
\end{equation}
\emph{Proof.} Suppose the compiled branch runs from a compiled input
\(\Sigma_{\mathsf{c}}\) that matches the source input \(\Sigma_{\mathsf{n}}\).
Consider the true-branch case; the false-branch case is symmetric. Formally,
assume:
\begin{equation}
\label{eq:app-val-branch-proof-assume}
\begin{aligned}
&\Sigma_{\mathsf{n}}\sim_{\mathsf{obs}}\Sigma_{\mathsf{c}},\\
&\llbracket \chi \rrbracket(\Sigma_{\mathsf{c}})=\mathsf{true},\\
&\langle W_{\mathsf{c},t},\Sigma_{\mathsf{c}}\rangle
\Downarrow_{\pi}\Sigma'_{\mathsf{c}}.
\end{aligned}
\end{equation}
The guard-consistency premise gives the same route on the source state:
\begin{equation}
\label{eq:app-val-branch-proof-guard}
\llbracket \chi \rrbracket(\Sigma_{\mathsf{n}})=\mathsf{true}.
\end{equation}
By
\(W_{\mathsf{c},t}\preceq_{\mathsf{obs}}W_{\mathsf{n},t}\) and
Equation~\ref{eq:app-val-branch-proof-assume}, there exists a true-case source
output \(\Sigma_t\) such that:
\begin{equation}
\label{eq:app-val-branch-proof-true-source}
\langle W_{\mathsf{n},t},\Sigma_{\mathsf{n}}\rangle
\Downarrow_{\pi}\Sigma_t
\quad\land\quad
\Sigma_t\sim_{\mathsf{obs}}\Sigma'_{\mathsf{c}}.
\end{equation}
Applying the source case-split condition from
Equation~\ref{eq:app-branch-source-split} to
Equations~\ref{eq:app-val-branch-proof-guard}
and~\ref{eq:app-val-branch-proof-true-source} gives a source output
\(\Sigma'_{\mathsf{n}}\) for the original region:
\begin{equation}
\label{eq:app-val-branch-proof-source-output}
\langle W_{\mathsf{n}},\Sigma_{\mathsf{n}}\rangle
\Downarrow_{\pi}\Sigma'_{\mathsf{n}}
\quad\land\quad
\Sigma'_{\mathsf{n}}\sim_{\mathsf{obs}}\Sigma_t.
\end{equation}
By transitivity of \(\sim_{\mathsf{obs}}\),
Equations~\ref{eq:app-val-branch-proof-true-source}
and~\ref{eq:app-val-branch-proof-source-output} imply
\(\Sigma'_{\mathsf{n}}\sim_{\mathsf{obs}}\Sigma'_{\mathsf{c}}\). Thus
\(\Sigma'_{\mathsf{n}}\) is the witness required by the refinement definition.
If the compiled guard evaluates to false, the same argument uses
\(W_{\mathsf{c},f}\preceq_{\mathsf{obs}}W_{\mathsf{n},f}\) and the false half
of \(\mathsf{Split}_{\chi}\). \hfill\(\square\)

\smallskip
\noindent\textbf{Loop.}
The loop induction step validates that a source region can be represented as
repeated execution of a body \(W_{\mathsf{n},b}\) under guard \(\chi\).
Instead of assuming that the source side already contains a structured
\(\mathsf{while}\), we define an operational iteration relation. The base case
performs zero iterations when the guard is false:
\begin{equation}
\label{eq:app-iter-zero}
\frac{
\llbracket \chi \rrbracket(\Sigma)=\mathsf{false}
}{
\mathsf{Iter}_{\chi}(W_b,\Sigma,\Sigma)
}
\tag{\textsc{Iter-Zero}}
\end{equation}
The step case performs one body dry run and then continues iterating:
\begin{equation}
\label{eq:app-iter-step}
\frac{
\begin{aligned}
&\llbracket \chi \rrbracket(\Sigma)=\mathsf{true}
\qquad
\langle W_b,\Sigma\rangle\Downarrow_{\pi}\Sigma_1
\\
&\mathsf{Iter}_{\chi}(W_b,\Sigma_1,\Sigma')
\end{aligned}
}{
\mathsf{Iter}_{\chi}(W_b,\Sigma,\Sigma')
}
\tag{\textsc{Iter-Step}}
\end{equation}
The source loop-decomposition premise then says that every behavior produced by
iterating the proposed source body is allowed by the original source region:
\begin{equation}
\label{eq:app-loop-source-decomposition}
\begin{aligned}
&\mathsf{LoopDecomp}_{\chi}(W_{\mathsf{n}},W_{\mathsf{n},b})\\
&\quad\coloneqq
\forall \Sigma,\Sigma'.\
\mathsf{Iter}_{\chi}(W_{\mathsf{n},b},\Sigma,\Sigma')
\Rightarrow\\
&\quad\quad
\exists \widehat{\Sigma}'.
\langle W_{\mathsf{n}},\Sigma\rangle
\Downarrow_{\pi}\widehat{\Sigma}'
\land
\widehat{\Sigma}'\sim_{\mathsf{obs}}\Sigma'.
\end{aligned}
\end{equation}
Under this loop-decomposition condition, a refinement of the loop body implies
that the compiled loop refines the original source region:
\begin{equation}
\label{eq:app-val-loop}
\frac{
\begin{aligned}
&W_{\mathsf{c},b}\preceq_{\mathsf{obs}}W_{\mathsf{n},b}
\\
&\mathsf{LoopDecomp}_{\chi}(W_{\mathsf{n}},W_{\mathsf{n},b})
\qquad
\mathsf{Consistent}_{\chi}
\end{aligned}
}{
\mathsf{while}\ \chi\ \mathsf{do}\ W_{\mathsf{c},b}
\preceq_{\mathsf{obs}}
W_{\mathsf{n}}
}
\tag{\textsc{Val-Loop}}
\end{equation}
\emph{Proof.} Suppose the compiled loop runs from a compiled input
\(\Sigma_{\mathsf{c}}\) that matches the source input \(\Sigma_{\mathsf{n}}\):
\begin{equation}
\label{eq:app-val-loop-proof-assume}
\begin{aligned}
&\Sigma_{\mathsf{n}}\sim_{\mathsf{obs}}\Sigma_{\mathsf{c}},\\
&\left\langle
\mathsf{while}\ \chi\ \mathsf{do}\ W_{\mathsf{c},b},
\Sigma_{\mathsf{c}}
\right\rangle
\Downarrow_{\pi}\Sigma'_{\mathsf{c}}.
\end{aligned}
\end{equation}
We first prove a stronger intermediate claim by induction on the compiled loop
execution: there exists a source-iteration result \(\Sigma_i\) such that
\begin{equation}
\label{eq:app-val-loop-proof-iter-claim}
\mathsf{Iter}_{\chi}(W_{\mathsf{n},b},\Sigma_{\mathsf{n}},\Sigma_i)
\quad\land\quad
\Sigma_i\sim_{\mathsf{obs}}\Sigma'_{\mathsf{c}}.
\end{equation}
For the zero-iteration case, the compiled guard is false and the compiled loop
returns the input state:
\begin{equation}
\label{eq:app-val-loop-proof-zero-compiled}
\llbracket\chi\rrbracket(\Sigma_{\mathsf{c}})=\mathsf{false}
\quad\land\quad
\Sigma'_{\mathsf{c}}=\Sigma_{\mathsf{c}}.
\end{equation}
By \(\mathsf{Consistent}_{\chi}\), the source guard is also false:
\begin{equation}
\label{eq:app-val-loop-proof-zero-source-guard}
\llbracket\chi\rrbracket(\Sigma_{\mathsf{n}})=\mathsf{false}.
\end{equation}
Therefore \textsc{Iter-Zero} gives
\(\mathsf{Iter}_{\chi}(W_{\mathsf{n},b},\Sigma_{\mathsf{n}},\Sigma_{\mathsf{n}})\).
Together with
\(\Sigma_{\mathsf{n}}\sim_{\mathsf{obs}}\Sigma_{\mathsf{c}}=\Sigma'_{\mathsf{c}}\),
this establishes Equation~\ref{eq:app-val-loop-proof-iter-claim} with
\(\Sigma_i=\Sigma_{\mathsf{n}}\).

For the step case, the compiled guard is true, the compiled body runs once, and
the compiled loop continues from the body output:
\begin{equation}
\label{eq:app-val-loop-proof-step-compiled}
\begin{aligned}
&\llbracket\chi\rrbracket(\Sigma_{\mathsf{c}})=\mathsf{true},\\
&\langle W_{\mathsf{c},b},\Sigma_{\mathsf{c}}\rangle
\Downarrow_{\pi}\Sigma_{\mathsf{c},1},\\
&\left\langle
\mathsf{while}\ \chi\ \mathsf{do}\ W_{\mathsf{c},b},
\Sigma_{\mathsf{c},1}
\right\rangle
\Downarrow_{\pi}\Sigma'_{\mathsf{c}}.
\end{aligned}
\end{equation}
By \(\mathsf{Consistent}_{\chi}\), the source guard is also true:
\begin{equation}
\label{eq:app-val-loop-proof-step-source-guard}
\llbracket\chi\rrbracket(\Sigma_{\mathsf{n}})=\mathsf{true}.
\end{equation}
By
\(W_{\mathsf{c},b}\preceq_{\mathsf{obs}}W_{\mathsf{n},b}\), there exists a
source body output \(\Sigma_1\) such that:
\begin{equation}
\label{eq:app-val-loop-proof-body-source}
\langle W_{\mathsf{n},b},\Sigma_{\mathsf{n}}\rangle
\Downarrow_{\pi}\Sigma_1
\quad\land\quad
\Sigma_1\sim_{\mathsf{obs}}\Sigma_{\mathsf{c},1}.
\end{equation}
Applying the induction hypothesis to the remaining compiled loop execution from
\(\Sigma_{\mathsf{c},1}\), using
\(\Sigma_1\sim_{\mathsf{obs}}\Sigma_{\mathsf{c},1}\), gives a state
\(\Sigma_i\) such that:
\begin{equation}
\label{eq:app-val-loop-proof-tail-iter}
\mathsf{Iter}_{\chi}(W_{\mathsf{n},b},\Sigma_1,\Sigma_i)
\quad\land\quad
\Sigma_i\sim_{\mathsf{obs}}\Sigma'_{\mathsf{c}}.
\end{equation}
Combining Equations~\ref{eq:app-val-loop-proof-step-source-guard},
\ref{eq:app-val-loop-proof-body-source}, and~\ref{eq:app-val-loop-proof-tail-iter}
with \textsc{Iter-Step} yields
\(\mathsf{Iter}_{\chi}(W_{\mathsf{n},b},\Sigma_{\mathsf{n}},\Sigma_i)\), which
establishes the strengthened claim.

Finally, apply the source loop-decomposition condition in
Equation~\ref{eq:app-loop-source-decomposition} to
Equation~\ref{eq:app-val-loop-proof-iter-claim}. There exists a source output
\(\Sigma'_{\mathsf{n}}\) such that:
\begin{equation}
\label{eq:app-val-loop-proof-source-output}
\langle W_{\mathsf{n}},\Sigma_{\mathsf{n}}\rangle
\Downarrow_{\pi}\Sigma'_{\mathsf{n}}
\quad\land\quad
\Sigma'_{\mathsf{n}}\sim_{\mathsf{obs}}\Sigma_i.
\end{equation}
By transitivity of \(\sim_{\mathsf{obs}}\),
Equations~\ref{eq:app-val-loop-proof-iter-claim}
and~\ref{eq:app-val-loop-proof-source-output} imply
\(\Sigma'_{\mathsf{n}}\sim_{\mathsf{obs}}\Sigma'_{\mathsf{c}}\). Thus
\(\Sigma'_{\mathsf{n}}\) is the witness required by the refinement definition.
\hfill\(\square\)

Together, \textsc{Val-Seq}, \textsc{Val-Branch}, and \textsc{Val-Loop} are the
induction steps: each checks that a natural-language region can be decomposed by
one workflow construct, and each composes recursively validated subworkflows
into a refinement of the original region.

\section{Implementation Details}
\label{app:implementation-details}

This appendix gives additional details on how the abstract artifact workflow
is lowered into the prototype runtime. These details are not part of the core
workflow language; they describe the concrete package that executes a compiled
workflow.

\begin{figure*}[t]
\centering
\includegraphics[width=\textwidth]{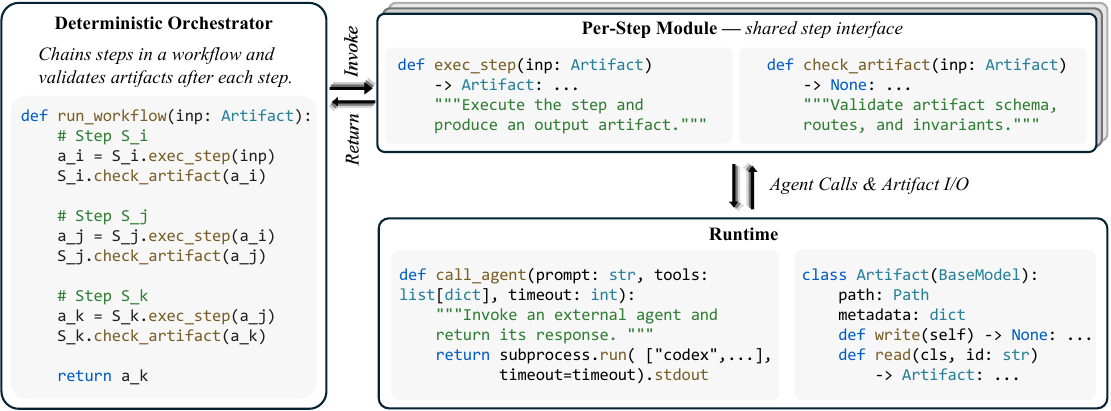}
\caption{Implementation structure of a compiled workflow package. The left
panel is the deterministic orchestrator, which invokes step modules and checks
their artifacts before continuing. The top-right panel shows the generated
per-step interface: \texttt{exec\_step} produces artifacts, while
\texttt{check\_artifact} enforces schema and quality gates. The bottom-right
panel is the runtime, which implements shared artifact I/O and external agent
calls. Arrows denote the orchestrator's invocation/return path and the step
module's runtime calls.}
\label{fig:execution-runtime-layers}
\end{figure*}

Figure~\ref{fig:execution-runtime-layers} shows the package structure. The
orchestrator is a deterministic driver. It invokes one step module at a time,
checks the artifacts produced by that step, and transfers control according to
the returned route. Each step module implements the shared step interface:
\texttt{exec\_step} transforms declared input artifacts into declared output
artifacts, and \texttt{check\_artifact} validates the produced artifact before
the orchestrator continues. The runtime provides the shared services used by
step modules, including artifact I/O, version metadata, and calls to external
agents or benchmark tools.

\noindent\textbf{Implementation lowering.}
Implementation lowering turns each workflow statement into a Python step
module. The compiler synthesizes a \texttt{exec\_step} function whose
signature matches the artifact-driven workflow interface. The function receives only the
artifacts declared as inputs and returns the declared output artifacts, a
next-step decision, and optional feedback for later steps. The body may combine
deterministic code with calls to benchmark tools or subagents, but these calls
must pass through the runtime and must write results to the declared artifact
paths.

The lowering validator checks the generated module against this interface. It
checks Python syntax, function signatures, declared output types, and routing
targets. It also uses program analysis and model-assisted prompt/entity
analysis to detect undeclared reads, such as a subagent prompt that mentions an
input field absent from the step interface.

\noindent\textbf{Runtime connection.}
Runtime connection keeps environment-specific code out of the generated step
modules. The runtime writes artifacts, retrieves the latest or earlier
versions, records metadata, and exposes concrete artifact paths to step code.
It also provides actor-call primitives for subagents, LLM checks, and benchmark
tools, so generated modules do not reimplement environment setup or artifact
passing.

Artifact versioning is the main runtime contract. Each artifact has a stable
name, while each committed value has a version and metadata. A step normally
reads the latest committed version of its input artifacts and writes a new
version for each output artifact. If a retry or loop produces a replacement,
the runtime preserves the previous version instead of overwriting it. This
lets diagnostics point to the exact artifact that failed a quality gate and
lets execution resume from the latest valid artifacts.

The runtime also implements the retry protocol. It runs a step, applies the
quality gate, and commits the artifacts only when the gate succeeds. If the
gate fails, the runtime records the feedback and invokes the step again with
the previous artifact path and the constraint feedback. After a step commits,
the orchestrator follows the generated control-transfer logic to choose the
next step. The runtime-level checks therefore enforce execution invariants:
artifacts are versioned consistently, failed attempts are not silently
committed, retry feedback is preserved, and every selected route exists in the
workflow package.

\section{Additional Evaluation Results}
\label{app:additional-evaluation-results}

\noindent\textbf{Evaluation models.}
Table~\ref{tab:evaluation-models} lists the compiler and executor models used
in the evaluation.

\begin{table}[t]
\centering
\scriptsize
\caption{Compiler and executor models used in evaluation. Model size is reported only when
it is public in the model name or provider-facing identifier.}
\label{tab:evaluation-models}
\begin{tabular*}{\columnwidth}{@{\extracolsep{\fill}}@{}lll@{}}
\toprule
Model & Role & Public size \\
\midrule
GPT-5.4 & Compiler & undisclosed \\
Sonnet-4.6 & Compiler & undisclosed \\
GLM-5 & Compiler/executor & 744B total / 40B active \\
GLM-4.7-Flash & Executor & 30B total / 3B active \\
GPT-OSS-120B & Executor & 120B \\
Qwen3-235B-A22B & Executor & 235B total / 22B active \\
Amazon Nova Micro & Executor & undisclosed \\
NVIDIA Nemotron Nano 3 & Executor & 30B total / 3.5B active \\
\bottomrule
\end{tabular*}
\end{table}

\noindent\textbf{Repeated-execution metric.}
\label{app:passk-details}
\passkmetric{} measures the probability that all \(k\) repeated executions of
the same instance succeed. In our repeated-run study, each instance is executed
ten times. If an instance succeeds in \(x\) of these ten executions, the
unbiased estimator for its \passkmetric{} value is
\(\binom{x}{k}/\binom{10}{k}\); averaging this quantity over instances gives
the solid curves in Figure~\ref{fig:chi-stability-passk}. We also report an
observed diagnostic: for each \(k\), we count the fraction of instances that
passed every observed repeat from 1 through \(k\). This observed curve is
ordering-dependent, while the unbiased estimator uses all ten repeats
symmetrically. The shaded range is \(\pm\) one standard deviation under an
independent-case plug-in variance estimate.

\begin{figure}[t]
\centering
\includegraphics[width=\columnwidth]{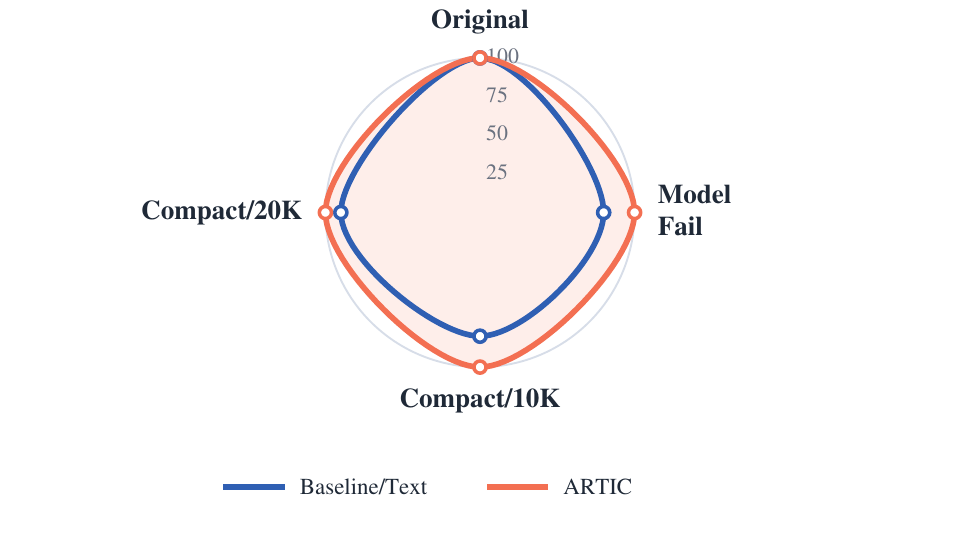}
\caption{Robustness comparison on the Medical domain in \(\chi\)-Bench. The
compiled workflow is evaluated against textual execution under controlled
perturbations: transient model request failure, context compaction at 10K
tokens, and context compaction at 20K tokens.}
\label{fig:chi-robustness-radar}
\end{figure}

\noindent\textbf{Environment perturbations.}
Figure~\ref{fig:chi-robustness-radar} summarizes the robustness study
discussed in Section~\ref{sec:evaluation}. We select cases where both the text
workflow and compiled workflow pass in the original setting, then introduce
environment perturbations such as transient model request failure and forced
context compaction. This setup measures whether a workflow that appears correct
under normal execution remains reliable when the executor faces common
operational frictions.

\begin{figure*}[t]
\centering
\subfloat[Input-token pressure.]{
  \includegraphics[width=0.9\columnwidth]{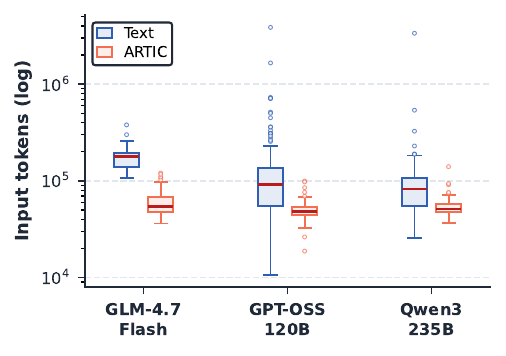}
  \label{fig:context-pressure-input}
}
\hfill
\subfloat[Output-token pressure.]{
  \includegraphics[width=0.9\columnwidth]{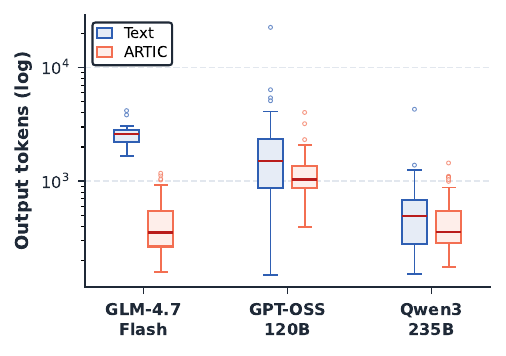}
  \label{fig:context-pressure-output}
}
\caption{Per-agent token pressure for text execution and compiled workflow
execution.}
\label{fig:context-pressure-token-plots}
\end{figure*}

\begin{table}[t]
\centering
\caption{Ablation of optimization, validation, and dry-run feedback on the
Medical domain in \(\chi\)-Bench using GLM-4.7-Flash.}
\label{tab:chi-ablation}
\begingroup
\newcommand{\cmark}{\checkmark}
\newcommand{\xmark}{\(\times\)}
\footnotesize
\renewcommand{\arraystretch}{1.08}
\begin{tabular}{cccc}
\toprule
Optimization & Validation & Dry-run & Performance \\
\midrule
\cmark & \cmark & \cmark & 88\% \\
\cmark & \cmark & \xmark & 80\% \\
\cmark & \xmark & \xmark & 72\% \\
\xmark & \cmark & \xmark & 48\% \\
\xmark & \xmark & \xmark & 0\% \\
\bottomrule
\end{tabular}
\endgroup
\end{table}

\begin{table}[t]
\centering
\caption{Compiler model comparison on the Medical domain in \(\chi\)-Bench under
GLM-4.7-Flash execution.}
\label{tab:compiler-model-comparison}
\begingroup
\footnotesize
\renewcommand{\arraystretch}{1.08}
\begin{tabular*}{\columnwidth}{@{\extracolsep{\fill}}lccc}
\toprule
Metric & GLM-5 & GPT-5.4 & Sonnet-4.6 \\
\midrule
Performance & 80.0\% & 76.0\% & 84.0\% \\
Workflow steps & 8 & 9 & 8 \\
Artifacts & 8 & 9 & 8 \\
Cost & \$2.02 & \$38.19 & \$13.27 \\
\bottomrule
\end{tabular*}
\endgroup
\end{table}

\begin{figure}[t]
\centering
\includegraphics[width=0.9\columnwidth]{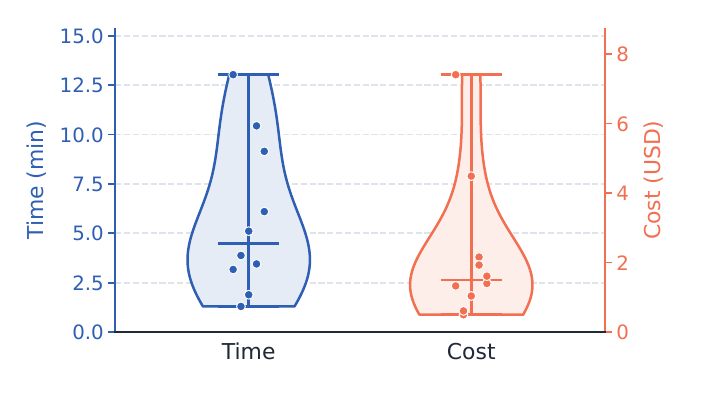}
\caption{First SOP-Bench compilation time and estimated model cost. The left
axis reports reconstructed compilation time in minutes, and the right axis
reports estimated GLM-5 compilation cost in USD.}
\label{fig:sop-compilation-time-cost}
\end{figure}

\noindent\textbf{Context pressure plots.}
Figure~\ref{fig:context-pressure-token-plots} reports the per-agent token
pressure behind the summary in Section~\ref{sec:evaluation}. Each point is one
executor turn: for textual execution, this is a turn in the single text-workflow
agent; for compiled execution, this is a turn in one of the subagents invoked by
the artifact workflow. The plot therefore measures local context burden rather
than total workflow cost.

\noindent\textbf{System-component ablation.}
Table~\ref{tab:chi-ablation} reports the component ablation study on the
Medical domain in \(\chi\)-Bench using GLM-4.7-Flash execution. Removing
dry-run feedback reduces pass rate from 88\% to 80\%; removing validation in
addition reduces it to 72\%; keeping validation but removing optimization
reduces it to 48\%; and removing all three leaves no passing cases. The
largest drop comes from removing optimization, because optimization formulates
where a prompt places too much context or decision burden on the executor.
Validation and dry-run feedback catch transformation errors that remain after
optimization. Table~\ref{tab:compiler-model-comparison} shows that the
compiler is also stable across compiler models: GLM-5, GPT-5.4, and
Sonnet-4.6 produce workflows with 80\%, 76\%, and 84\% pass rates,
respectively, compared with 48\% for the text workflow under the same executor
setting.

\noindent\textbf{Compilation overhead.}
Figure~\ref{fig:sop-compilation-time-cost} summarizes the time and estimated
model cost of the first package-producing compilation run for each retained
SOP-Bench domain. The estimates use the reconstructed compilation-time and
prompt/response-character cost records described in the data manifest; they
exclude later targeted repair runs.

\end{document}